\documentclass[useAMS,usenatbib]{mn2e}

\makeatletter
\def\thebibliography#1{\section*{REFERENCES}
 \addcontentsline{toc}{section}{REFERENCES}
 \list{}{\usecounter{dummy}%
         \labelwidth 0pt
         \leftmargin 1.5em
         \itemsep 0pt
         \itemindent-\leftmargin}
 \reset@font\small
 \parindent 0pt
 \parskip 0pt plus .1pt\relax
 \def\newblock{\hskip .11em plus .33em minus .07em}
 \sloppy\clubpenalty4000\widowpenalty4000
 \sfcode`\.=1000\relax
}
\def\here{\global\advance\count\@currbox -1}
\makeatother

\usepackage[T1]{fontenc}
\usepackage{ae,aecompl}

\usepackage{amsmath}    
\usepackage{color}
\usepackage{graphicx}\graphicspath{{plots/}}
\usepackage{epstopdf}
\usepackage{balance}

\newcommand{\eprint}[2][]{{\tt\if!#1!#2\else#1:#2\fi}}

\usepackage[latin1]{inputenc}
\newcommand{\ds}{\hphantom{0}} 
\newcommand{\ps}{\hphantom{.}} 
\newcommand{\os}{\hphantom{/}} 
\DeclareInputText{183}{\ds}    
\DeclareInputText{172}{\ps}    

\DeclareMathSymbol{:}{\mathord}{operators}{"3A}

\newcommand{\dg}{^{\circ}\kern-0.1em}
\newcommand{\Msun}{M_\odot}

\newcommand{\rhodm}{\rho_\mathrm{dm}}
\newcommand{\Mdm}{M_\mathrm{dm}}
\newcommand{\Mgas}{M_\mathrm{gas}}
\newcommand{\Mtot}{M_\mathrm{tot}}
\newcommand{\Mstar}{M_\mathrm{\star}}
\newcommand{\Mstargas}{M_\mathrm{\star/gas}}

\newcommand{\Sigmastargas}{\Sigma_\mathrm{\star/gas}}

\newcommand{\Nbar}{N_\mathrm{bar}}
\newcommand{\epsilonbar}{\epsilon_\mathrm{bar}}
\newcommand{\epsilondm}{\epsilon_\mathrm{dm}}
\newcommand{\Ndm}{N_\mathrm{dm}}
\newcommand{\Nbulge}{N_\mathrm{bulge}}
\newcommand{\Ndisk}{N_\mathrm{disc}}
\newcommand{\Ngas}{N_\mathrm{gas}}
\newcommand{\rpolar}{r_\mathrm{polar}}
\newcommand{\ld}{l_\mathrm{d}}
\newcommand{\fb}{f_\mathrm{b}}
\newcommand{\fdisk}{f_\mathrm{d}}
\newcommand{\fgas}{f_\mathrm{gas}}
\newcommand{\fBH}{f_\mathrm{BH}}
\newcommand{\mbar}{m_\mathrm{bar}}
\newcommand{\mdm}{m_\mathrm{dm}}
\newcommand{\dsep}{d_\mathrm{sep}}
\newcommand{\vvir}{v_\mathrm{vir}}
\newcommand{\rhalf}{r_{1/2}}
\newcommand{\Rhalf}{R_{1/2}}
\newcommand{\dhalf}{d_{1/2}}
\newcommand{\Msat}{M_\mathrm{S}}
\newcommand{\Mhost}{M_\mathrm{H}}
\newcommand{\rsat}{r_\mathrm{S}}
\newcommand{\rJac}{r_\mathrm{J}}

\newcommand{\ditto}{{\tiny $'\kern-0.2em'$}}

\usepackage{tabularx}
\newcommand{\topline}{\hline\\[-7pt]}
\newcommand{\midline}{\\[3pt]\hline\\[-7pt]}
\newcommand{\dblline}{\\[3pt]\hline\hline\\[-7pt]}
\newcommand{\botline}{\\[3pt]\hline}

\title[Outer Stellar Halos]{The Outer Stellar Halos of Galaxies:\\
how Radial Merger Mass Deposition, Shells and Streams
depend on Infall-Orbit Configurations}

\author[G. S. Karademir et al.]{Geray S. Karademir$^{1}$\thanks{E-mail: karademir@usm.lmu.de},
Rhea-Silvia Remus$^{1,2}$,
Andreas Burkert$^{1,3}$,
Klaus Dolag$^{1,4}$,
\newauthor Tadziu L. Hoffmann$^{1}$,
Benjamin P. Moster$^{1,4}$,
Ulrich P. Steinwandel$^{1,4}$,
and
\newauthor Jielai Zhang$^{2,5,6,7}$
\vspace{6pt}\\
$^{1}$Universit\"ats-Sternwarte M\"unchen, Scheinerstra{\ss}e~1, 81679 M\"unchen, Germany\\
$^{2}$Canadian Institute for Theoretical Astrophysics, 60 St. George Street, University of Toronto, Toronto ON M5S 3H8, Canada\\
$^{3}$Max Planck Institute for Extraterrestial Physics, Giessenbachstr. 1, D-85748 Garching, Germany\\
$^{4}$Max Planck Institute for Astrophysics, Karl-Schwarzschild-Str. 1, D-85741 Garching, Germany\\
$^{5}$Department of Astronomy and Astrophysics, University of Toronto, Canada\\
$^{6}$Dunlap Institute for Astronomy and Astrophysics, Canada \\
$^{7}$Schmidt Science Fellows in Partnership with the Rhodes Trust }

\date{Accepted 2019 April 30. Received 2019 April 30; in original form 2018 August 30} 

\pubyear{2019} 

\begin{document}
\label{firstpage}
\pagerange{\pageref{firstpage}--\pageref{lastpage}}
\maketitle

\begin{abstract}

Galaxy mergers are a fundamental part of galaxy evolution.
To study the resulting mass distributions of different kinds of
galaxy mergers, we present a simulation suite of 36~high-resolution
isolated merger simulations, exploring a wide range of parameter
space in terms of mass ratios ($\mu=1:5$, $1:10$, $1:50$, $1:100$)
and orbital parameters.
We find that mini mergers deposit a higher fraction of their mass in
the outer halo compared to minor mergers, while their contribution
to the central mass distribution is highly dependent on the orbital
impact parameter:
for larger pericentric distances we find that the centre of the host
galaxy is almost not contaminated by merger particles.
We also find that the median of the resulting radial mass distribution
for mini mergers differs significantly from the predictions of simple
theoretical tidal-force models.
Furthermore, we find that mini mergers can increase the size of
the host disc significantly without changing the global shape of
the galaxy, if the impact occurs in the disc plane, thus providing
a possible explanation for extended low-surface brightness disks
reported in observations.
Finally, we find clear evidence that streams are a strong indication of
nearly circular infall of a satellite (with large angular momentum),
whereas the appearance of shells clearly points to (nearly) radial
satellite infall.

\end{abstract}

\begin{keywords}
galaxies:~halos -- galaxies:~dynamics -- galaxies:~interactions --
galaxies:~structure -- methods:~numerical\vspace{-5.8pt}
\end{keywords}


\section{Introduction}
\label{sec:Intro}

In the standard cosmological picture, galaxy mergers are a basic and
significant process in galaxy evolution.
Mergers are important for the hierarchical growth of galaxies, their
morphological appearance, their kinematic evolution, and many more
aspects \citep{NaabOstriker2017}.
In our current understanding, the major contributor to stellar mass
growth are major mergers ($\mu > 1/4$), while minor ($1/10 < \mu <
1/4$) and mini mergers ($\mu < 1/10$) contribute roughly the same
amount of mass to the final galaxy \citep{Gomez2017}.
However, regardless of the mass growth, \citet{Hilz2012} showed that,
in the case of spheroidal galaxy merger events, minor mergers lead
to an increased size growth and a more rapid surface density profile
shape change compared to major mergers.
In the case of dry (i.e., gas-poor) minor mergers they leave the
inner host structure almost unchanged \citep{Hilz2013}.

This is in agreement with \citet{Lagos2018}, who found that dry mergers
increase the stellar mass density in the outskirts of the galaxy,
while wet mergers tend to build up the bulge of the host galaxy.
Using isolated merger simulations, \citet{Amorisco2017} showed that
more massive mergers deposit their stars further inside the host galaxy
than smaller galaxies, and more concentrated merging satellites can
deposit mass further inside than their less compact counterparts of
the same mass.

Manifestations of these accretion processes in the form of tidal
streams and shells are observed in up to $26\%$ of the galaxies
\citep{Atkinson2013}, with massive galaxies with $M>10^{10.5}\Msun$
being more likely to exhibit tidal features.
Similarly, \citet{Hood2018} report a fraction of 17\% exhibiting shells
or tidal streams from the RESOLVE survey, inspecting more than 1000
galaxies, finding gas-rich galaxies to have a larger likelihood to
show signs of tidal interactions than gas-poor galaxies.
\citet{Delgado2010} also report that several spiral galaxies in the
local Universe show significant numbers of giant stellar structures
which are remnants of an earlier accretion or interaction process.
In case of early-type galaxies alone, \citet{Duc2017} reports a
fraction of about 40\%.
In order to improve the uncertainty of by-eye identification, a fully
automated method for identifying and classifying substructures was
developed recently by \cite{Hendel2018} (see also \cite{Canas2019}
for a new approach on identifying stellar halos and structures in
simulations).

\citet{Johnston2008} already showed that the appearance of shells and
streams is a good tracer for the merger history of galaxies, as they
strongly depend on accretion time and angular momentum properties of
the merger event.
Recently, \citet{Pop2017} have shown from hydrodynamical cosmological
simulations that shells appear around 18\% of their massive galaxies
and are usually made through merger events with mass ratios between
$1:1$ and $1:10$ by satellites that merge on low angular momentum
orbits.
Similarly, \citet{Amorisco2015} showed that shells usually result
from more radial merger orbits, while streams are made from merging
orbits that are more circular.

The disruption of (smaller) merging galaxies is a plausible
mechanism for the origin of diffuse stellar halo light (DSHL) on
all scales, i.e., the component of the outer stellar halo that does
not show structures like streams and shells anymore\footnote{It is
possible that these structures are still present but below the
detection limits.}
\citep[e.g.,][]{Zolotov2009,Cooper2013,Cooper2015,Longobardi2018,Hartke2018}.
Also, the characteristics of the stellar halo age distribution of the
Milky-Way are best reproduced as being formed by the accretion of small
satellite galaxies of ${\sim}\,10^{9.5}\Msun$ \citep{Carollo2018}.
For galaxy clusters, the DSHL is well studied in the form of the
intra-cluster light \citep[e.g.,][]{Mihos2005,Alamo2017,Montes2018},
and cosmological simulations of galaxy clusters have shown that merger
events are the key contributors for the build-up of the diffuse
stellar component in these clusters \citep[see][and references
therein]{Murante2007}.
For galaxies, the DSHL is dominated by the disruption of satellites
with a stellar mass of ${\sim}\,10^{8.5}\Msun$, and its properties
are defined by the corresponding merger history \citep{Purcell2007}.
It can extend up to a few hundreds of $\mathrm{kpc}$
\citep{Zolotov2009}.

This diffuse DSHL (as well as other stellar features in the outskirts
of galaxies indicating recent merger events) has recently come back
into focus of galaxy research with big surveys for both early-type
(e.g., MATLAS \citep{Duc2015} and VEGAS \citep{Spavone2017}) and
late-type galaxies (e.g., Dragonfly \citep{Merritt2016} and GHOSTS
\citep{Radburn2011,Monachesi2016}), finding extended stellar halos
around most galaxies, although some late-type galaxies have been found
to show no or barely any signs of an extended stellar halo component
\citep{Merritt2016, Streich2016}.
Using full cosmological simulations, \citet{Pillepich2014} and
\citet{Remus2017} showed that the shape of the diffuse outer stellar
halo correlates with the accretion history and the total mass of
a galaxy, thus providing insight into the details of a galaxy's
formation.

In this work we perform isolated binary merger simulations with
extremely high resolution, testing a large parameter space in orbital
configurations and mass ratios, to understand the dependence of the
mass deposition on these configurations in more detail.
As it is already well known from previous work that major merger
events completely change the appearance of the galaxies by mixing
the stellar (and gaseous) components of both galaxies completely,
we do not include these major mergers in our study but instead focus
on the impact of minor~($1:4>\mu>1:10$) and mini~($\mu<1:10$) mergers.
Especially, we concentrate on understanding the build-up of the
stellar halo component without the disruption of an existing disc
component in the host, to better understand the large diversity of
halos observed around disc galaxies.
In Section~\ref{sec:method} we describe the merger simulations we
perform in detail, and in Section~\ref{sec:Results} we analyse the
dependence of the radial mass deposition of the mergers on the orbital
configurations and the merger mass ratio.
Finally, in Section~\ref{sec:discgrowth} we investigate the impact of
the merger event on the disc structure of the host and the appearance
of shells and streams.
We summarize and discuss our results in Section~\ref{sec:Discussion}.

\section{The Simulation Sample}
\label{sec:method}

All simulations were carried out using the Tree-SPH code GADGET-3,
which is itself descended from the publicly available code GADGET-2
\citep{Springel2005}.
We used a state-of-the-art SPH implementation as presented by
\citet{Beck2016} to overcome known weaknesses of the SPH formalism that
have been pointed out over the last decade (see \citealt{Agertz2007}
for further details).
We used time-dependent artificial viscosity and conduction to enhance
the properties of SPH in terms of shock capturing and particle noise
reduction in post-shock regions as well as on the surfaces of contact
discontinuities.

Moreover, the simulations include complex physics like radiative
cooling for optically thin gas which is in ionized equilibrium
with a specified background of ultraviolet (UV) radiation
\citep{Katz1996} and star formation \citep{SpringelHernquist2003}
using a sub-resolution stochastic star-formation approach, where one
stellar particle represents a statistical ensemble of stars following
the \cite{Salpeter1955} initial mass function (IMF).
Supernova feedback from the stars interacts with the interstellar
medium (ISM).
Also, a supermassive black hole (SMBH) was placed in the center of
each galaxy, which can accrete gas with a Bondi-Hoyle accretion model
with a hard Eddington limit, and is able to re-emit thermal feedback
into its surroundings, as described by \cite{Johansson2009}.
However, these effects are only of minor importance for the main
objective of the current work, which is concerned with the final mass
distribution of the remnant.

\subsection{Galaxy set-up}
\label{subsec:gals}

The initial conditions for the simulation runs were created using
the method described by \citet{Springel2005a}.
We set up the isolated disc galaxies to consist of a dark matter halo,
a spherical stellar bulge, and an exponential stellar and gas disc.

The dark matter halo is modeled as a spherical Hernquist profile
\citep{Hernquist1990, Hernquist1993}, given by
\begin{align}
\rhodm(r) = \frac{\Mdm}{2\pi}\frac{a}{r(r+a)^3}.
\end{align}
Here, $\Mdm$ is the total dark matter mass of the halo, and $a$
is the scaling parameter.
The Hernquist profile is chosen over the NFW profile
\citep{Navarro1997} due to its advantageous analytical properties,
in particular that its mass is finite.
The scaling parameter $a$ for all our simulations is chosen such
that the Hernquist profile in the center matches an NFW profile
with a concentration parameter of $c=12$, using Equation~2 from
\citet{Springel2005a}.

Similarly, the bulge component is also modeled as a Hernquist profile,
with a mass fraction of $\fb=0.01367$ with respect to the total mass
of the galaxy.

The stellar disc and the gas disc are set up with an exponential
surface density profile given as
\begin{align}
\Sigmastargas =
\frac{\Mstargas}{2\pi\ld^2}
\exp\left(-\frac{\rpolar}{\ld}\right),
\end{align}
with $\Mstar$ the stellar mass and $\Mgas$ the gas mass in the disc,
$\ld$ a scaling parameter equal for both discs, and $\rpolar$ the
radial component in cylindrical coordinates.
The exponential disc contains a gas fraction of $\fgas=0.2$ and the
total mass fraction of both components in the disc is $\fdisk=0.041$
with respect to the total mass of the galaxy.
The softening-length for the baryonic particles of the simulations
is $\epsilonbar=0.02~\mathrm{kpc}$, while for the dark matter halo
$\epsilondm=0.083~\mathrm{kpc}$.

The galaxies also contain a black hole in their centers, with a mass
fraction of $\fBH=1.186\times10^{-5}$ of the total mass.
This results in a mass of $\mbar \sim 1.88\times10^4\Msun$ for
each baryonic particle and $\mdm \sim 1.88\times10^{5}\Msun$
for the dark matter particles, which gives us sufficient mass
resolution to study the merger process in great detail.

\begin{table}
\vspace{-.3\baselineskip}
\centering
\caption{Main parameters of the created galaxies.}
\label{tab:galsparam}
\begin{tabular}{lcccc} \topline
  & Mass  & $v_{200}$ & $\Mtot$          & $\Nbar$ \\
  & ratio & (km/s)    & ($10^{10}\Msun$) & \dblline
host      & 1\os\ds\ds\ds &    160 &      188.7 &      4\,000\,000 \midline
satellite & 1/5\ds\ds     & \ds 94 &   \ds 37.7 &    \ds\,800\,000 \\
          & 1/10\ds       & \ds 74 &   \ds 18.9 &    \ds\,400\,000 \\
          & 1/50\ds       & \ds 43 & \ds\ds 3.8 & \ds\,\ds 80\,000 \\
          & 1/100         & \ds 34 & \ds\ds 1.9 & \ds\,\ds 40\,000 \botline
\end{tabular}
\vspace{\baselineskip}
\end{table}

For this study, five disc galaxies are created using the method
described above.
Our main galaxy, hereafter referred to as the host
galaxy, is a Milky-Way like galaxy with a total mass of
$\Mtot=1.89\times10^{12}\Msun$, corresponding to a virial
velocity of $v_{200}=160~\mathrm{km/s}$.
The galaxy consists of 10.8~million particles in total, with
$\Ndm=6.8\times10^6$~dark matter halo particles,
$\Nbulge=1\times10^6$~bulge stellar particles,
$\Ndisk=2.4\times10^6$~disc stellar particles, and
$\Ngas=0.6\times10^6$~gas particles.

Because we are interested in seeing how non-spheroidal, two-component
galaxies with a disc and a bulge (instead of only simple spheroidal
dwarf galaxies) behave in the mergers, we set up the merging galaxies,
hereafter referred to as satellite galaxies, as downscaled copies of
our host galaxy.
Simply downscaling the galaxies is somewhat of an idealization,
but it was motivated by the need to provide intrinsically stable
satellite galaxies for the simulations.
The resulting main parameters of the host galaxy and the four
merging satellite galaxies (downscaled in mass by factors of $1/5$,
$1/10$, $1/50$, and $1/100$ from the host galaxy) are summarized in
Table~\ref{tab:galsparam}.

\subsection{Merger set-up}
\label{subsec:merger}

We performed simulations with eight different orbital parameter
configurations, where each configuration was performed for four
different mass ratios ($\mu=1:5$, $1:10$, $1:50$, $1:100$), leading
to a total of 32 individual galaxy merger simulations.
In addition, we performed four simulations with merger mass ratios
of $\mu=1:100$ with special orbital configurations.
The resulting 36 simulations cover a broad range of different orbital
parameters as well as different mass ratios.
As the mass deposition in major mergers has already been studied
extensively in previous works, here we focus solely on the contribution
of minor mergers ($\mu=1:5$, $1:10$) and mini mergers ($\mu=1:50$,
$1:100$).

\begin{figure}
\centerline{\includegraphics[width=1.0\columnwidth]{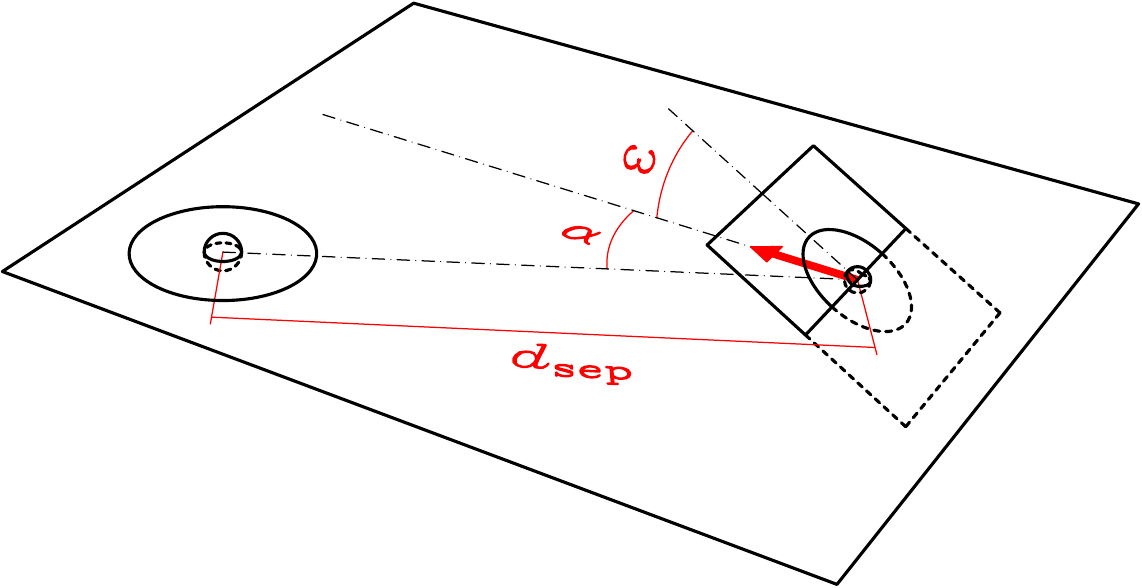}}
\caption{The geometrical set-up for the Family~1 mergers.
In this family the satellite approaches the host in the plane of
the host disc, with the impact parameter determined by the angle
$\alpha$, while the disc inclination angle of the satellite galaxy
is kept fixed at $\omega = 30\dg$.
The initial distance for all mergers is $\dsep=80~\mathrm{kpc}$.
The resulting pericentric distances for the different values of
$\alpha$ are listed in Table~\ref{tab:pericentric}.}
\label{fig:fam1}
\end{figure}

\begin{table}
\vspace{-\baselineskip}
\centering
\caption{Pericentric distances for the Family~1 mergers.}
\label{tab:pericentric}
\begin{tabular}{lcccc} \topline
impact parameter $\alpha$  & $10\dg$ & $20\dg$ & $30\dg$ & $40\dg$ \midline
pericentric distance (kpc) & 1.8     & 7.0     & 15.0    & 24.8    \botline
\end{tabular}
\vspace{\baselineskip}
\end{table}

\begin{figure}
\centerline{\includegraphics[width=0.8\columnwidth]{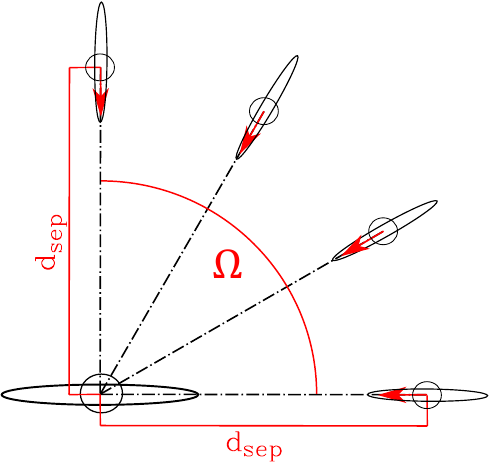}}
\caption{The geometrical set-up for the Family~2 mergers.
In this family, the impact parameter is kept fixed at $\alpha=0\dg$,
while the satellite galaxy approaches the host galaxy from a varying
angle $\Omega$ above the disc plane of the host galaxy.
The initial distance for all mergers is $\dsep=80~\mathrm{kpc}$.}
\label{fig:fam2}
\end{figure}

For the host galaxy, we use the orientation of the disc of this galaxy
as the main plane ($x$--$y$ plane) instead of the orbital plane,
and describe the inclination of the orbital plane of the merger using
the angle $\Omega$ as shown in Fig.~\ref{fig:fam2}.
The impact parameter of the collision is defined by the angle
$\alpha$, which is an equivalent for the pericentric distance, with
the corresponding pericentric distances to each value of $\alpha$
used in this study listed in Table~\ref{tab:pericentric}.
These parameters are chosen because they provide freedom in modifying
the parameters in every conceivable way, always using the disc plane
of the host galaxy as the frame of reference (to better allow studying
the impact of the mergers on the host disc).

For all simulations the initial distance between the galaxies is set
to $\dsep=80~\mathrm{kpc}$.\footnote{In our case this is a reasonable
distance in a physical sense because it is large enough to keep the
initial interaction between the galaxies at a minimum, while at the
same time being small enough to avoid wasting computational time on
the initial approach.}
The initial approach velocity is equal to the virial velocity regarding
the host galaxy, which yields to $\vvir\approx144.1~\mathrm{km/s}$.
The velocity vector is initially at an angle $\alpha$ to the
geometrical line connecting the two galaxies, as indicated in
Fig.~\ref{fig:fam1}.

Each simulation is run for approximately $3.5~\mathrm{Gyr}$ after
the merger event, which we define at the timestep the discs touch,
with two simulation runs extended to $5.5~\mathrm{Gyr}$ after the
merger event to study the long term effects.

Our set of simulations can be split into two families.
In Family~1, the merger orbits are always in the plane of
the host-galaxy disc, i.e., $\Omega=0\dg$, while the merging
galaxy is tilted by an angle of $\omega=30\dg$ relative to this
plane\footnote{This arbitrary value has been chosen to make the orbit
more realistic.}.
The impact parameter is varied, $\alpha\in\{10\dg,20\dg,30\dg,40\dg\}$,
which basically means that the orbits change from almost radial to
more circular.
A sketch of the orbit setup is shown in Fig.~\ref{fig:fam1}.

For Family~2, the mergers are always performed on radial orbits, i.e.,
$\alpha=0\dg$, and the disc inclination for the satellite galaxy is
fixed at $\omega=0\dg$.
The orbital angle is varied from edge-on to the host galaxy
($\Omega=0\dg$) to face-on ($\Omega=90\dg$).
A sketch of the geometrical setup for Family~2 is shown in
Fig.~\ref{fig:fam2}.

The four special case simulations are cross-combinations between
the systematic variations of Family~1 and Family~2 with the aim
to probe the impact of a combination of the impact angle $\alpha$
and the orbital angle $\Omega$, and have been performed only for the
$\mu=1:100$ merger mass ratios since these are the low-cost runs in
our simulation set-up.
Here we have kept the disc inclination angle of the satellite galaxy
for all simulations at $\omega=0\dg$.
The parameter combinations we have chosen are:
$\alpha=0\dg$, $\Omega=10\dg$; $\alpha=10\dg$, $\Omega=10\dg$; and
$\alpha=10\dg$, $\Omega=30\dg$.
The last simulation in the special case family also has an orbital
configuration of $\alpha=10\dg$ and $\Omega=10\dg$ identical to the
second simulation, but in this simulation the disc rotation of the
satellite is set up to be retrograde to the disc rotation of the host.

The parameters for all simulations performed in this study are listed
in Table~\ref{tab:orbitparam}.

\begin{table}
\vspace{-.3\baselineskip}
\centering
\caption{Orbital parameters of all mergers.}
\label{tab:orbitparam}
\begin{tabular}{l|ccccc} \topline
& $\alpha$ & $\Omega$ & $\omega$ & Mass ratios & Remarks \dblline
Family 1 &    $10\dg$ & $\ds 0\dg$ &    $30\dg$ & all & --- \\
         &    $20\dg$ &   \ditto   &   \ditto   &     & --- \\
         &    $30\dg$ &   \ditto   &   \ditto   &     & --- \\
         &    $40\dg$ &   \ditto   &   \ditto   &     & --- \midline
Family 2 & $\ds 0\dg$ & $\ds 0\dg$ & $\ds 0\dg$ & all & --- \\
         &   \ditto   &    $30\dg$ &   \ditto   &     & --- \\
         &   \ditto   &    $60\dg$ &   \ditto   &     & --- \\
         &   \ditto   &    $90\dg$ &   \ditto   &     & --- \midline
Special  & $\ds 0\dg$ &    $10\dg$ & $\ds 0\dg$ & only $1:100$ & --- \\
Cases    &    $10\dg$ &    $10\dg$ &   \ditto   &     & --- \\
         &    $10\dg$ &    $30\dg$ &   \ditto   &     & --- \\
         &    $10\dg$ &    $10\dg$ &   \ditto   &     & retrograde \botline
\end{tabular}
\vspace{\baselineskip}
\end{table}

\section{Radial mass deposition}
\label{sec:Results}

One of the major questions in observations and simulations of stellar
halos around galaxies is how to decipher the merger history of a
galaxy from only one snapshot in time.
For this task, understanding the details of where mass from a disrupted
satellite is deposited is of key importance.
In order to study the mass deposition of the merger we calculated
the corresponding surface density distributions of each remnant
approximately $3.5~\mathrm{Gyr}$ after the merger event.

\subsection{Orbital and mass-ratio dependence}
\label{subsec:Mass-depo}

\begin{figure*}
\centerline{\includegraphics[width=\textwidth]{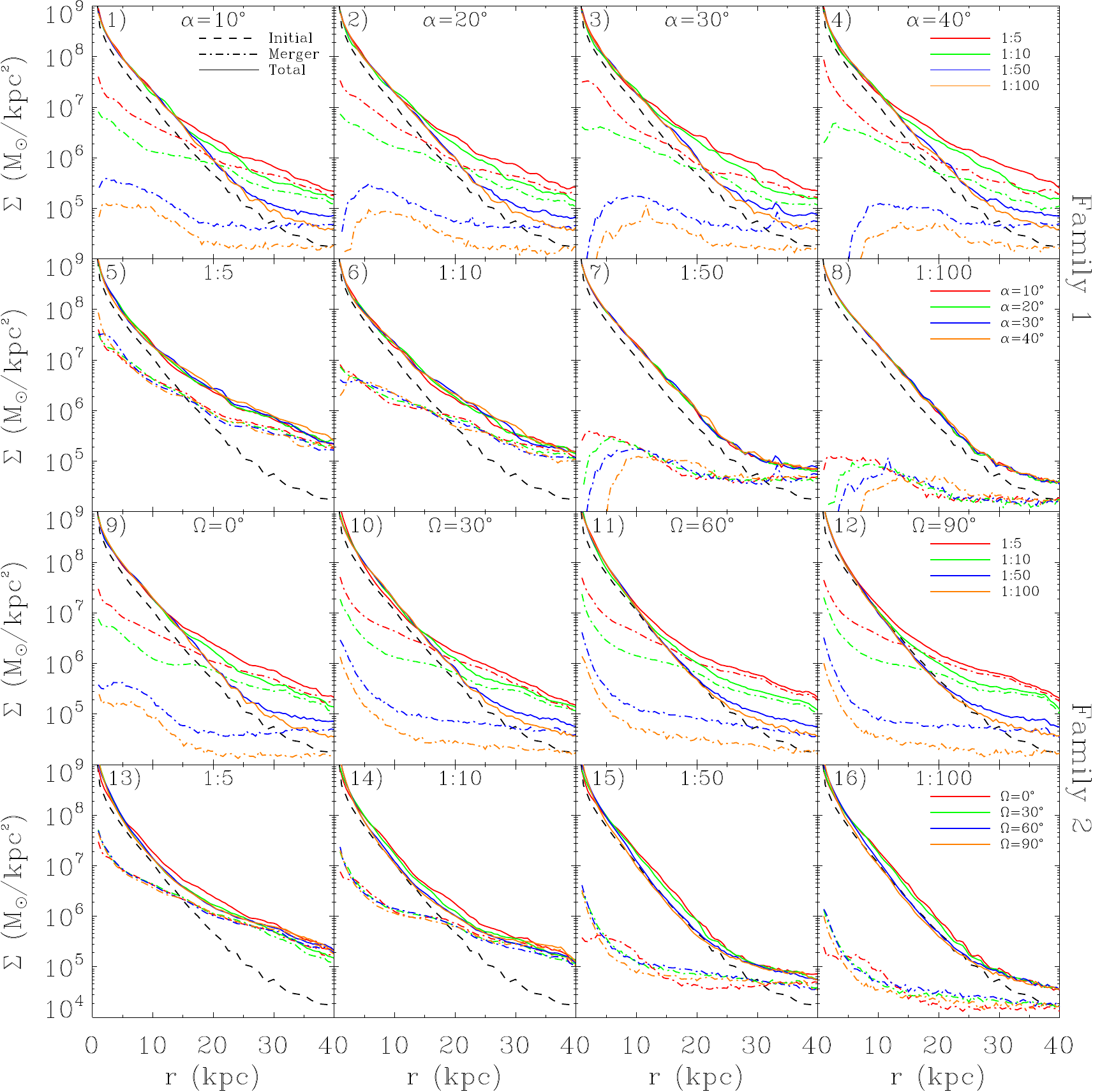}}
\caption{Radial surface density distribution for all simulations at
${\sim}\,3.5~\mathrm{Gyr}$ after the beginning of the merging.
Solid lines show the total profiles, while dash-dotted profiles mark
the radial contribution from the smaller merging galaxy.
In all panels, the radial surface density profile of the undisturbed
initial main disc galaxy is shown in black.
The upper two rows belong to Family~1, the lower two rows to Family~2.
The first row compares all mass ratios at a given $\alpha$, with
$\alpha=10\dg$, $\alpha=20\dg$, $\alpha=30\dg$, and $\alpha=40\dg$
from left to right, while different colors mark the different mass
ratios in each plot of this row.
The second row compares results for a fixed mass ratio of $1:5$,
$1:10$, $1:50$, and $1:100$ from left to right, while different colors
mark the different values of $\alpha$.
The third row shows all mass ratios at a fixed $\Omega$, with
$\Omega=0\dg$, $\Omega=0\dg$, $\Omega=0\dg$, and $\Omega=0\dg$ from
left to right, with colors again indicating different mass ratios as
for the uppermost row.
In the last row results for different $\Omega$ at a fixed mass ratio
are shown in each panel, with the mass ratios ordered as in the
second row and the different colors indicating the different values
of $\Omega$ used for the mergers.
\vspace{1\baselineskip}}
\label{fig:massdepo}
\end{figure*}

The radial surface density profiles for all merger simulations from
Family~1 and Family~2 at ${\sim}\,3.5~\mathrm{Gyr}$ after the merger
event are shown in Fig.~\ref{fig:massdepo}.
The first two rows show the results from Family~1, the last two rows
show the results from Family~2.
In the first and third row, the mergers of different mass ratios at
a given impact angle $\alpha$ (first row) or orbital angle $\Omega$
(third row) are compared to each other, as indicated in the individual
panels.
In the second and last row, the mergers of a fixed mass ratio with
varying $\alpha$ (second row) and $\Omega$ (last row) are compared.
Colored solid lines show the resulting surface density profiles for
the complete stellar component of the remnant, while the dashed colored
lines mark the individual contributions of the satellite galaxies.
For reference, the dashed black line in each panel shows the surface
density of the isolated, undisturbed host galaxy.\footnote{We let
the disc galaxy evolve for the same length as the merger simulations,
to account for effects from numerical heating.}

One of the global trends concerning the surface density deposition
that can be immediately seen in all panels of Fig.~\ref{fig:massdepo}
is that, at a certain radius, the contribution from the accreted
satellite galaxy starts to dominate the final surface density profile,
while the inner part is always dominated by the contribution from
the host galaxy.
This behavior is independent of the impact angle $\alpha$ or the
orbital angle $\Omega$.
This crossover point depends mainly on the mass ratio of the merger:
a more massive merger has a crossover point closer to the center
of the remnant than a less massive merger, where the crossover
point lies at larger radii (see panels 1 to 4 and panels 9 to 12 in
Fig.~\ref{fig:massdepo}).
The total mass deposited inside the remnant's disc range depends
mainly on the mass ratio:
while a $1:100$ merger is only able to deposit ${\sim}\,25\%$ of its
initial baryonic mass inside the disc range of the remnant, a $1:5$
merger can deposit up to $80\%$.
This is in agreement with the results by \citet{Amorisco2017} and
confirms earlier studies by \citet{Hilz2013}, who showed that minor
mergers result in faster size growth, while major mergers result in
higher mass growth.

It can also be seen globally that the mini mergers only have a slight
impact on the radial surface density distribution of the merger
remnant, compared to the initial galaxy.
For mini mergers, the surface density profiles of the merger remnant
follow the surface density profile of the initial disc galaxy out to
large radii of up to ${\sim}\,30~\mathrm{kpc}$ (${\sim}\,4\,\rhalf$),
while minor mergers already have a significant influence on the surface
density profiles at ${\sim}\,15~\mathrm{kpc}$ (${\sim}\,2\,\rhalf$).
For all mergers, the enrichment of the outskirts is much more
significant than that of the center of the galaxy.

Starting with Family~1, for the minor mergers ($1:5$ and $1:10$)
the surface density distribution is very similar for different impact
angles $\alpha$ (panels~5 and 6 of Fig.~\ref{fig:massdepo}).
Mergers of both mass ratios are able to reach the center of the
remnant, and therefore significantly disturb the host galaxy.

For the mini mergers ($1:50$ and $1:100$), we find that the impact
angle $\alpha$ significantly influences the stellar deposition range
of the satellite:
for large impact parameters, the satellite cannot contribute
to the remnant galaxy's center anymore (panels 7 and 8 of
Fig.~\ref{fig:massdepo}).
Instead, the innermost ${\sim}\,5$ to $20~\mathrm{kpc}$
(${\sim}\,1\dots 3\,\rhalf$) are devoid of any contribution from the
satellite galaxy, depending on the mass ratio and impact parameters.
The shape of the radial distribution, however, is similar for a given
mass ratio, and mainly shifts to the outer regions with increasing
$\alpha$, thereby leaving an increasingly larger region in the center
almost vacant.

For Family~2, the mass deposition behaves similar to Family~1 in
some aspects.
The mass deposition for the minor mergers (panels~13 and 14 of
Fig.~\ref{fig:massdepo}) again only varies slightly for different
inclination angles $\Omega$, with no clear systematic trends.
Like in Family~1, the center of the host galaxy is always influenced
by the merger event.

For the mini mergers, the variation of the inclination angle is much
more significant.
If $\Omega \geq 30\dg$, the distribution varies only marginally
(panels 15 and 16 of Fig.~\ref{fig:massdepo}), but for smaller $\Omega$
it shows the same behavior as for Family~1, with the central region
becoming devoid of stars from the accreted satellite galaxy.
For the mass deposition of a satellite galaxy it is therefore important
whether the satellite has to travel through the stellar disc of the
host or not.

\begin{figure}
\centerline{\includegraphics[width=1.0\columnwidth]{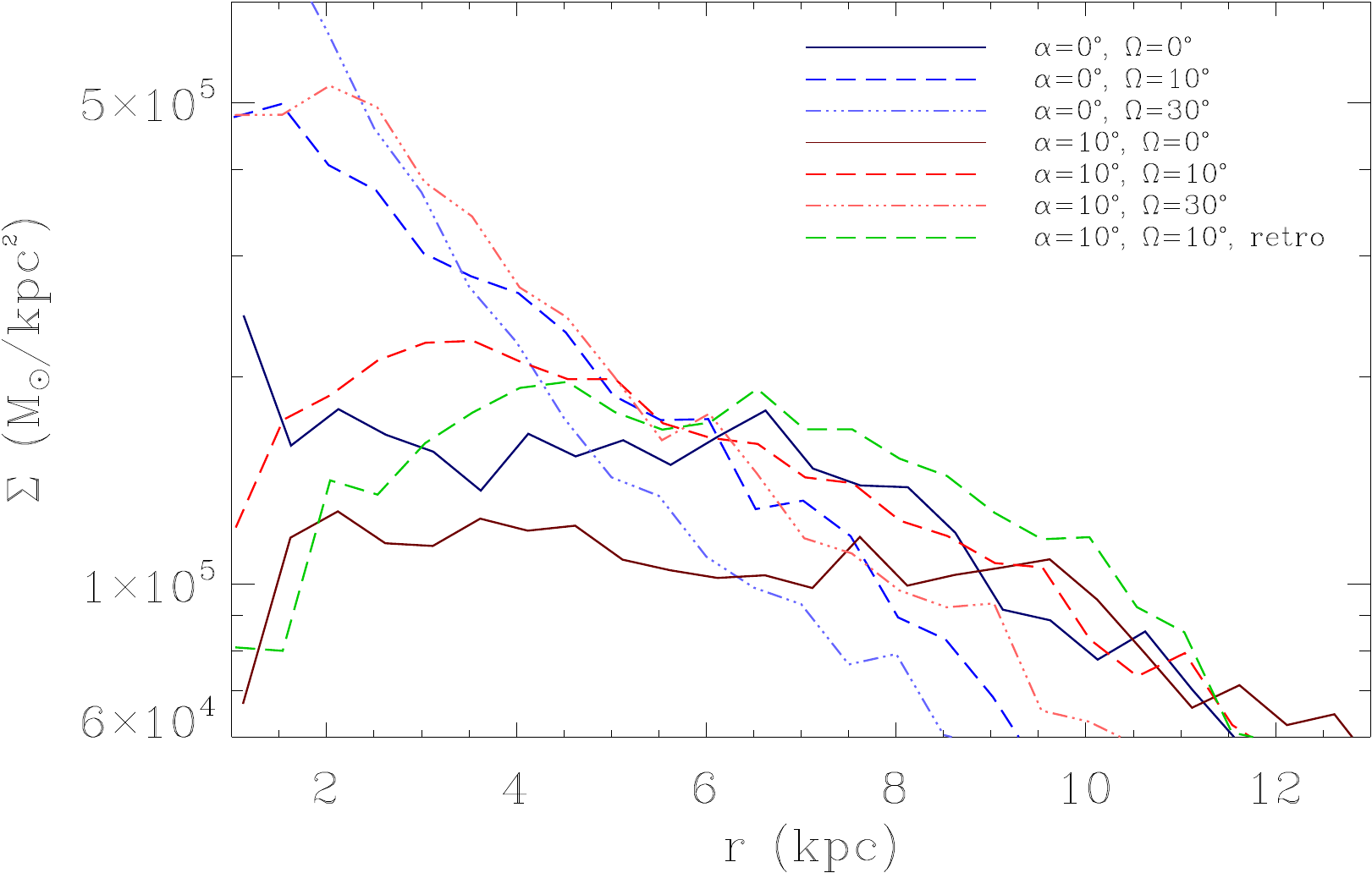}}
\caption{Central surface density distributions of the former
satellite particles for seven different $\mu=1:100$ mergers with
orbital configurations as indicated.
Here the different effects of the parameters $\alpha$ and $\Omega$
can be seen:
an increase in $\alpha$ shifts the distribution outwards, while an
increase in $\Omega$ works in the opposite direction and increases
the density in the central region.}
\label{fig:specialcases}
\end{figure}

Since the distribution of the orbital angles of (mini) mergers is
not uniform but rather has its highest probability at $\Omega=0\dg$
(i.e., along the disc plane of the host; \citealt{Shao2018, Welker2017,
Welker2018}), we additionally examine the impact of combinations of
small values of $\alpha$ and $\Omega$ for our $\mu=1:100$ mergers.
The resulting central radial surface brightness distributions of
the former satellite particles of these mini mergers are shown in
Fig.~\ref{fig:specialcases}.
For $\alpha=0\dg$ and $\Omega\geq 30\dg$, the density increases
steadily inwards.
For lower $\Omega$ and larger $\alpha$, the density in the center
decreases, and for $\alpha\geq 20\dg$, $\Omega=0\dg$ a region with
lower density appears in the center (see the Family-1 mini mergers
with $\mu=1:100$ in panel~8 of Fig.~\ref{fig:massdepo}).
The two parameters $\alpha$ and $\Omega$ work opposite to each other.
While an increase of $\Omega$ increases the density in the central
region, a larger $\alpha$ decreases it and shifts the distribution
outwards.
For the mergers shown in Fig.~\ref{fig:specialcases}, the fraction of
the total satellite mass deposited in the radial range shown (i.e.,
inside of $13~\mathrm{kpc}$) varies only slightly between $11\dots
15\%$, without any distinctive correlation with the orbital parameters.

In addition, one counter-rotating merger simulation was performed
(green dashed line in Fig.~\ref{fig:specialcases}).
It can be seen that an inversion of the rotation results in a density
distribution shifted outwards but otherwise similar to its co-rotating
counterpart (red dashed line).
At large radii, however, there are no differences in the mass
deposition of the mini mergers, regardless of the orbital
configurations.
We conclude that the effect of the orbit and impact angles can play an
important role for the stellar halos of disc galaxies, as the survival
of the disc galaxy can crucially depend on these orbital parameters.

\subsection{Resulting S\'ersic-Indices}
\label{subsec:sersic}

As discussed in Section~\ref{subsec:Mass-depo}, mergers can strongly
influence the resulting surface density profile shape of the remnant,
depending on the mass ratio and impact parameters.
In order to give a rough classification of the remnant, we
study the light distribution of the remnant (assuming a constant
mass-to-light ratio) and calculate the corresponding S\'ersic-Index~$n$
\citep{Sersic68} parametrizing the radial surface brightness profile
\begin{align}
\ln(I_R)=\ln(I_0)-kR^{1/n}
\end{align}
for each remnant from a face-on surface brightness distribution within
a radius of $80~\mathrm{kpc}$.

We find that for all mergers of both families, the variation of the
S\'ersic-index due to the merger event follows the same behavior:
while the initial galaxy has an index of $n\approx 1.3$ (as expected
for an idealized exponential late-type galaxy), it varies after the
collision, mainly depending on the different mass ratios. 
It has to be noted here that newly formed stars (either from the gas in the host or the merger) only make a small contribution.
For the $\mu=1:5$ merger, the remnants show indices of $n\approx
3.5\dots 6.5$, corresponding to a typical value for an early-type
galaxy.
The index of the remnant decreases with the mass of the satellite:
the $1:10$ mergers result in $n\approx 2.2\dots 4.5$, spanning the
whole spectrum of S\'ersic indices from late-type to early-type
galaxies.
It has to be noted that by increasing $\Omega$ larger S\'ersic indices
are reached, while a change in $\alpha$ only results in small changes
around $n\approx 2.2$.
This clearly shows that minor mergers are fully capable of changing the
morphology of a disc galaxy into an S0 or even an early-type galaxy,
but for smaller satellites the orbital configuration of the merger
is of increasing importance for initiating a morphological change.

For a mass ratio of $1:50$, the resulting index reaches $n\approx
1.3\dots 2.0$, and for the $1:100$ mergers we find $n\approx 1.3\dots
1.4$, almost identical to the index of the initial host galaxy.
We conclude that mini mergers are not able to significantly disturb
the host disc galaxy; thus, mini mergers appear to be viable candidates
for building up stellar halos around disc galaxies.

Interestingly, the orbit for mini mergers plays almost no role in
establishing the S\'ersic index, while for minor mergers the impact
of the orbits on the profile is of much larger importance and the
spread in $n$ is much larger.
This spread is due to the orbital parameter $\Omega$, with lower
$\Omega$ causing smaller $n$.
This result can be explained by the larger influence of
particle-particle interactions if the satellite is infalling on an
orbit in the plane of the host disc.

\subsection{Mass distribution by component}
\label{subsec:parttype}

\begin{figure}
\centerline{\includegraphics[width=1.0\columnwidth]{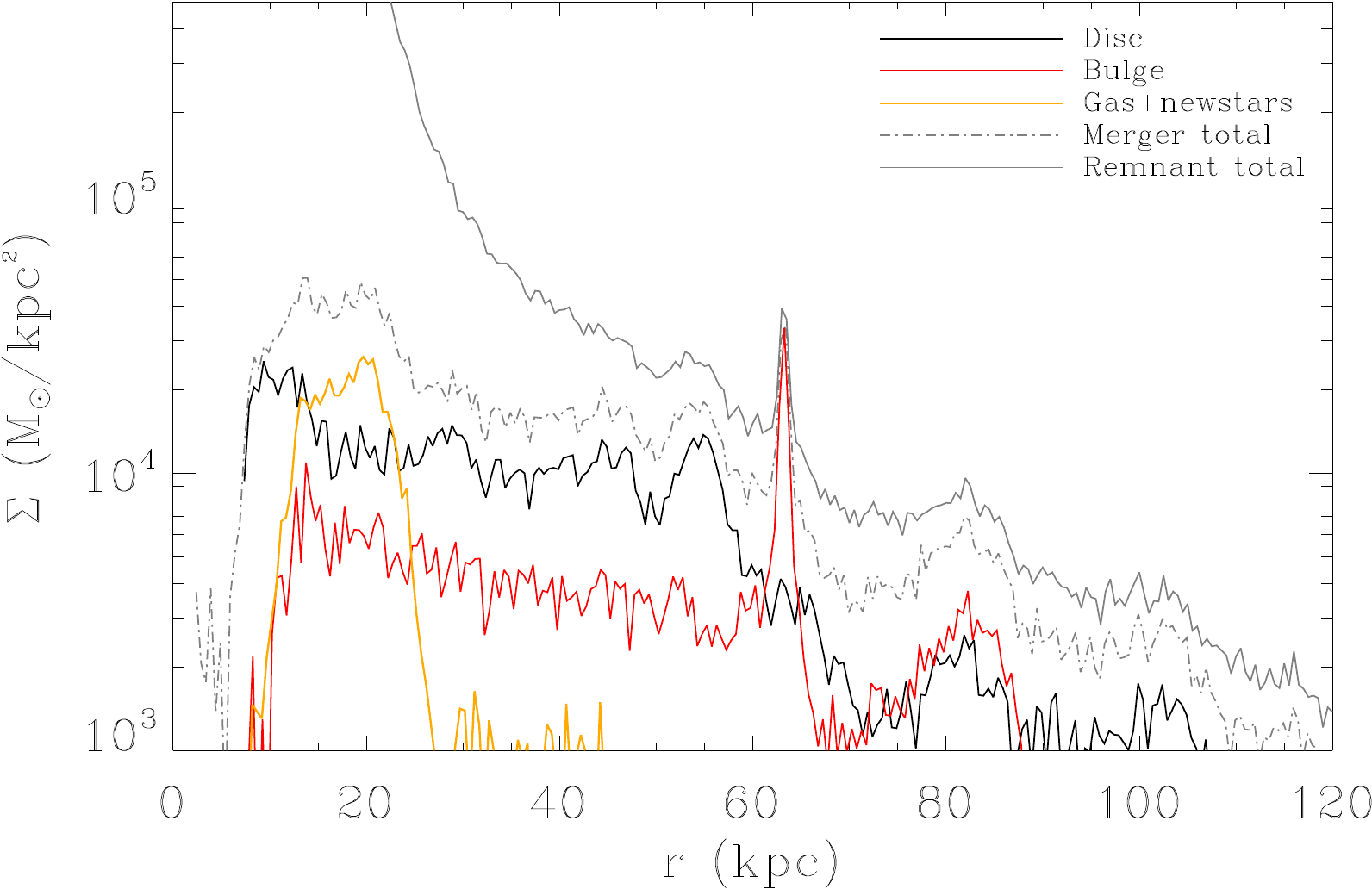}}
\caption{Radial surface density for the Family~1 mini merger
($\mu=1:100$) with $\alpha=40\dg$, split up into the different
components of the satellite:
the contributions from the disc and bulge components of the accreted
satellite are shown as solid black and red lines, respectively.
The yellow solid line indicates the contribution of newly formed
stars and the gas component of the satellite galaxy, while the full
contribution from the accreted component is shown as a dash-dotted
gray line.
For comparison, the solid gray line shows the total radial surface
density profile of the remnant.}
\label{fig:parttype}
\end{figure}

For our merger simulations, we used satellite galaxies that were
downscaled versions of Milky Way-like disc galaxies, i.e., our
satellites have separate stellar disc and bulge components and, due
to the initial gas component in the disc, also a component of newly
formed stars.
Thus, we can investigate the contribution to the mass deposition from
the satellite's components at different radii.

As expected, the gas component settles into the central regions of
the remnant in all cases.
For mini mergers, this is at the position of the innermost peak,
which is comprised largely of gas and newly formed stars in addition
to former disc and bulge particles, which can be clearly seen in
Fig.~\ref{fig:parttype}.
This is in agreement with the picture of the gas of the satellite
transferring angular momentum to the particles of the host, as a result
of which the gas particles can sink down to the central regions of
the remnant.

We also find that the contribution of former disc particles to the
center of the remnant is higher than that of the bulge particles due to
the higher amount of mass contained in the disc compared to the bulge.
But especially in the outer regions, basically no bulge particles
remain, only disc particles.
This is due to the disc stars being the most loosely bound component
of the satellite, and therefore large parts of the disc get stripped
with high angular momentum.
An additional feature which is illustrated in Fig.~\ref{fig:parttype}
is the peak in the bulge particle distribution at $r=64~\mathrm{kpc}$.
This peak consists of the surviving core of the merger, which is
still orbiting the remnant as a bound accumulation of stars, losing
mass with every orbit.

\subsection{A simple theoretical model}

To better understand the characteristics of the distribution of
the stripped satellite particles, we compare this with a strongly
abstracted theoretical model.
We assume a satellite with a Hernquist density profile with initial
mass $\Msat$ and half-mass radius $\rhalf$ that loses its outer, least
bound particles as a result of the interaction with the potential
of the host galaxy, which we also take to have a Hernquist density
profile, with mass $\Mhost$ and half-mass radius $\Rhalf$.

We consider two different descriptions for the conditions under which
a particle will be stripped from the satellite.
The first one (model~``A'') is the conventional case of disruption
via the tidal forces from the host potential.
If we define the satellite radius $\rsat$ as the radius that encloses
the remaining mass $m$, which for a Hernquist profile is
\begin{equation}
\rsat(m) = \frac{a}{\displaystyle\sqrt{\frac{\Msat}{m}}-1},
\end{equation}
where $a=\rhalf/(1+\sqrt{2})$ is the initial scale radius, then the
criterion for stripping the outer particles of the satellite by the
tidal forces of the host at distance $d$ from its center is
\begin{equation}
\frac{\Mhost(d)}{d^2}-\frac{\Mhost(d+\rsat)}{(d+\rsat)^2}=\frac{m}{\rsat^2},
\label{eqn:tidal}
\end{equation}
where $\Mhost(r)=\Mhost\,r^2/(r+a)^2$ is the cumulative mass of the
Hernquist-profile host.
We divide the satellite into 1000 equal-mass shells, and solve
Eq.~\ref{eqn:tidal} numerically to obtain the deposition distance $d$
for each mass shell.
The simplifying assumption here is that the entire mass shell is
deposited where particles at the satellite edge facing the host center
become unbound, disregarding internal kinematics of host and satellite,
and ignoring any internal rearrangement of the satellite as a result
of mass loss.

The basic premise of model~A is that the candidate particle and the
satellite core are both (apart from their own interaction) acted upon
only by the gravitational force of the host galaxy.
But this assumption may not be entirely valid:
the tighter bound core of the satellite behaves as a joint massive
particle inside the host and is thus subject to additional dissipative
forces such as dynamical friction, which do not affect a stripped
particle.
In such a situation the candidate particle will remain with the
satellite core only if the binding force between particle and satellite
core is stronger than the gravitational force between particle and
host galaxy; if not, the particle will become separated from the
satellite core and follow an increasingly different trajectory.

\begin{figure}
\centerline{\includegraphics[width=1.0\columnwidth]{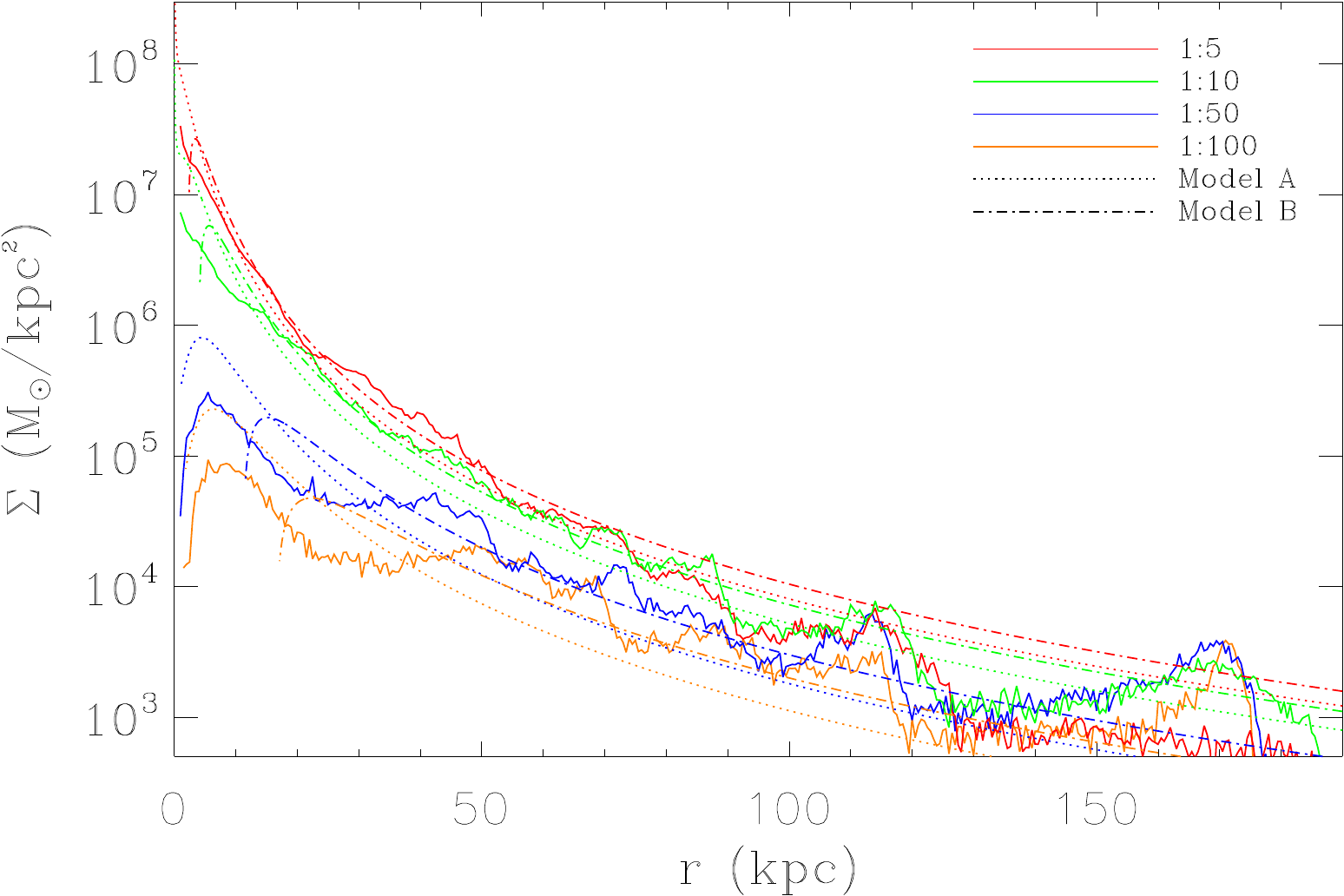}}
\caption{Comparison of the post-merger surface brightness distributions
of the former satellite stars for all mass ratios in the Family~1
mergers with $\alpha=20\dg$ (solid lines) with those predicted by our
simple theoretical models A (dotted lines) and B (dash-dotted lines),
scaled to the same merger mass ratios.}
\label{fig:theorydata}
\end{figure}

To take this consideration into account, in our model~B we therefore
use the corresponding criterion
\begin{equation}
\frac{\Mhost(d)}{d^2}=\frac{m}{\rsat^2}
\label{eqn:potential}
\end{equation}
as the condition for the removal of particles from the satellite at
distance $d$ from the host center.
Since in this model the satellite core is not assumed to be
free-falling with the particle, the condition for the particle to
become unbound will already occur earlier (i.e., farther away from
the host).
As before, we divide the satellite into 1000 equal-mass shells,
and solve Eq.~\ref{eqn:potential} for $d$ to obtain the deposition
distance for each mass shell.
The same simplifying assumptions are made in model~B as in model~A
(i.e., internal kinematics of host and satellite are disregarded,
and internal rearrangement of the satellite as a result of mass loss
is ignored).

Despite the crude approximations, both of our simple models display
many of the characteristics of the resulting distributions from
the simulation.
As can be seen in Fig.~\ref{fig:theorydata}, both models reproduce
the global shape of the resulting distribution quite well.
Model~A in particular reproduces a specific feature of the simulations,
namely the decrease of the surface brightness in the central regions
for the mini mergers ($1:100$ and $1:50$) in conjunction with the
absence of such a decrease for the minor mergers ($1:10$ and $1:5$).
However, model~A over-predicts the amount of mass in the central
regions in all cases;
model~B shows a better average agreement with the simulations, in
particular for the mini mergers, but it under-predicts the central
density.
As the central regions accumulate a larger fraction of newly
formed stars and gas from the satellite, which the simple model
is not designed to consider, this latter result may not be entirely
unsurprising, and model~B appears to offer a better overall description
of the distribution of the satellite particles in the remnant.

Two related effects contribute to the differences between the simple
theoretical models and the simulation.
First, all the mass in the theoretical model satellites is contained
in a Hernquist sphere, i.e., the model satellites are much more
concentrated than the satellites of the simulations, where only the
bulge stars are described by a Hernquist sphere, but not the disc.
Therefore, the idealized model satellite would contribute more to the
center of the remnant than a less concentrated satellite of the same
mass, as shown by \citet{Amorisco2017}.
Second, the theoretical model assumes that stripping particles from
the satellite does not influence the inner regions of the satellite.
But this would not be the case for a sphere with an isotropic velocity
distribution, since the sphere is only dynamically stable if there is
an equilibrium between inward- and outward-moving particles, which,
however, will be disturbed by stripping.
Thus, the remaining part of the satellite in our theoretical model
stays more tightly bound than in a more realistic description.

\begin{figure}
\centerline{\includegraphics[width=1.0\columnwidth]{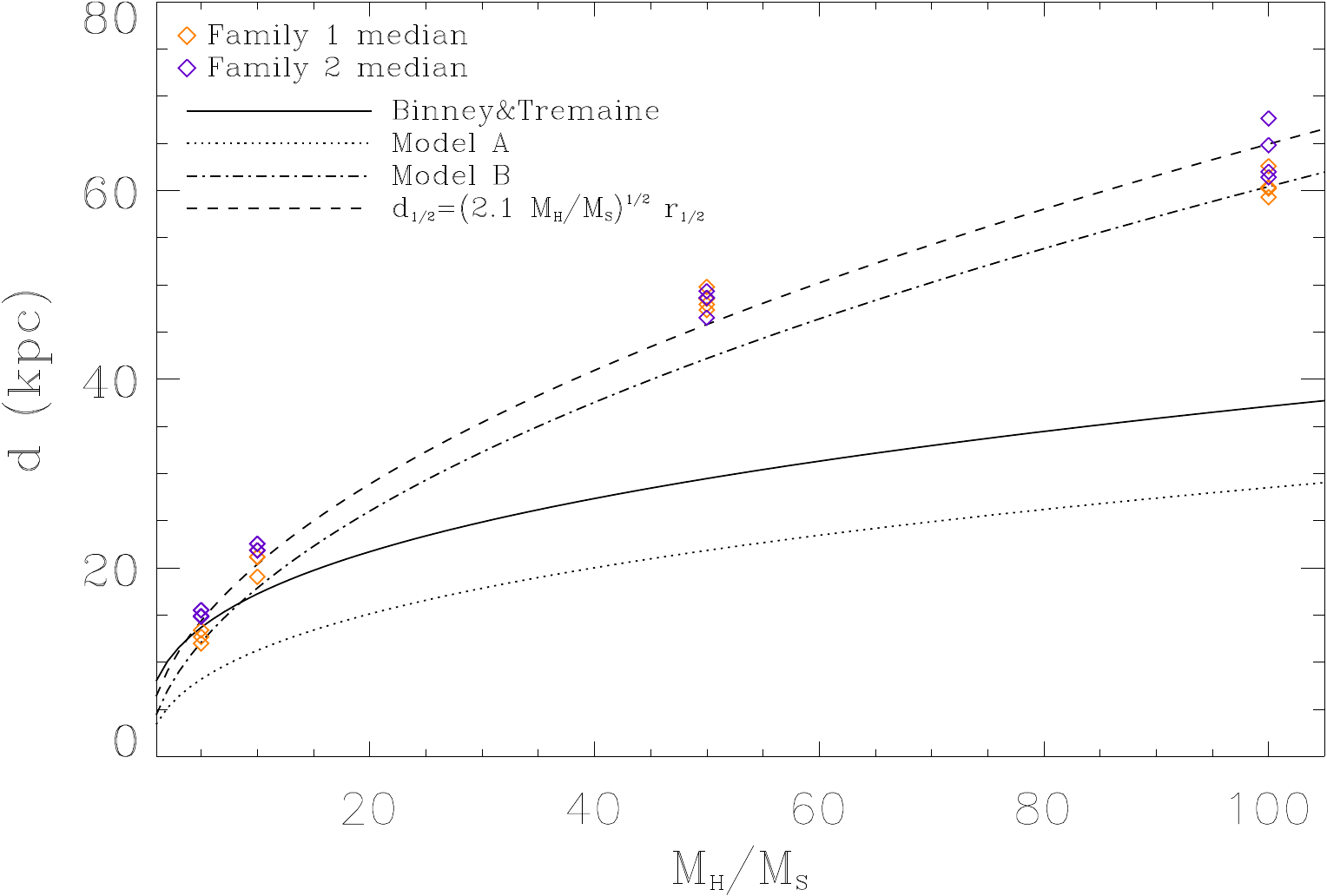}}
\caption{The median of the radial mass distribution of the former
satellite stellar particles for all Family~1 (orange diamonds) and
Family~2 (violet diamonds) mergers, compared to the corresponding
medians predicted by idealized theoretical models.}
\label{fig:median}
\end{figure}

A different way to quantify the behavior observed in the simulations
is to examine the medians of the radial mass distributions found for
the satellite galaxy particles in our different merger simulations.
With respect to our theoretical models, this median corresponds to
the distance from the host at which half of the satellite's particles
have been stripped as it moves inwards.
This distance can also be expressed via the tidal or Jacobi radius
$\rJac$ in a two-body system, given by \cite{BinneyTremaine} as
\begin{align}
\rJac\approx\left(\frac{m}{3M}\right)^{1/3}R_0,
\label{eqn:jacobi}
\end{align}
where $m$ and $M$ are the masses of the orbiting bodies and $R_0$
their separation.
For our purposes, $\rJac=\rhalf$, $m=\Msat/2$, $M=\Mhost$, and
$R_0=\dhalf$, and therefore
\begin{align}
\dhalf=\rhalf\left(\frac{6\Mhost}{\Msat}\right)^{1/3}.
\label{eqn:jacobimedian}
\end{align}
In Figure~\ref{fig:median} we plot the medians of the radial mass
distributions of the satellite particles from the simulations, compared
to this analytical approximation and the corresponding results from
our simple theoretical models A and B.

The curve resulting from our model~A (dotted line) has a shape
very similar to that of the analytic curve using the Jacobi radius
(solid line).
Both of these describe the stripping purely as a result of tidal
forces, our model~A assuming a satellite in free fall towards the host,
the analytic description assuming a satellite on a circular orbit.
In our Eq.~\ref{eqn:tidal}, however, $d$ is the distance between
the host center and the leading edge of the satellite, whereas in
Eq.~\ref{eqn:jacobimedian} it refers to the distance between host
center and satellite center, and thus our model~A curve lies at least
one half-mass radius further inwards than the analytic curve.
Furthermore, our model assumes the host to be a Hernquist sphere,
which has a shallower potential gradient (and thus allows the satellite
to penetrate further into the host before being stripped) than the
point-mass assumed for the host in the analytic description.

The curve resulting from our model~B (dash-dotted line)%
\begin{align}
\dhalf=\rhalf\left(\sqrt{\frac{2\Mhost}{\Msat}}
                  -\frac{\Rhalf/\rhalf}{1+\sqrt{2}}\right)
\end{align}
(plotted for $\Rhalf/\rhalf=1$) shows a much better match to the
actual medians from the simulations, for which we have sketched a
rough fit as the dashed line.
This result indicates that a description based solely on tidal forces
does not capture all essential aspects of an actually occurring merger.
Of course, one aspect that our simple model cannot describe is the
tendency of a disturbed system to continuously re-virialize, leading
to an expansion of the infalling satellite simultaneously to the
stripping process.
This would lead to a larger characteristic radius, lower densities,
and an easier stripping of the remaining mass, overall resulting in
mass being deposited at larger radii.
This effect has been recently described by \cite{Bosch2018} for
dark-matter-only halos, and is likely to occur for galaxies as well.

\section{Shells, Streams, and Disk Growth}
\label{sec:discgrowth}

\begin{figure}
\centerline{\includegraphics[width=1.0\columnwidth]{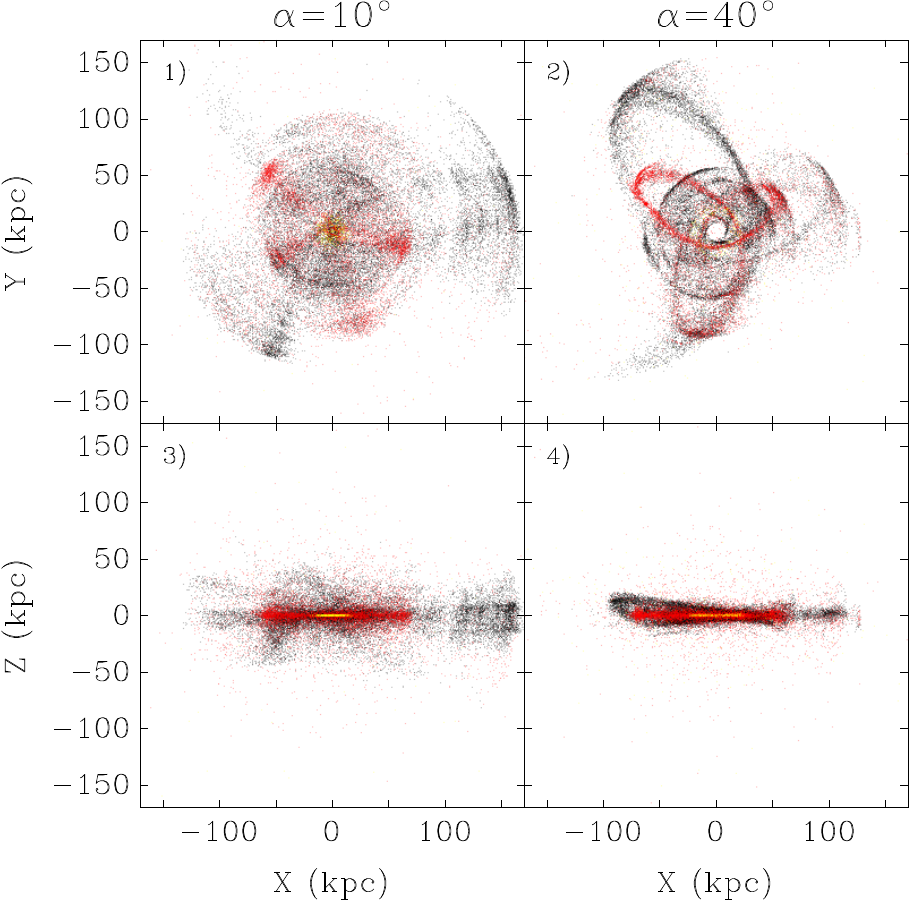}}
\caption{Shells and streams for two examples of $\mu=1:100$ mergers
at $t=5.5~\mathrm{Gyr}$ after the merger event.
For the simulations, all parameters are identical but the impact angle
$\alpha$, with $\alpha=10\dg$ for the example in the left panels and
$\alpha=40\dg$ for the example in the right panels.
Only the stars from the satellite galaxy are shown here, with disc
stars in black, bulge stars in red, and newly formed stars in yellow.}
\label{fig:shell}
\end{figure}

\begin{figure}
\centerline{\includegraphics[width=\columnwidth]{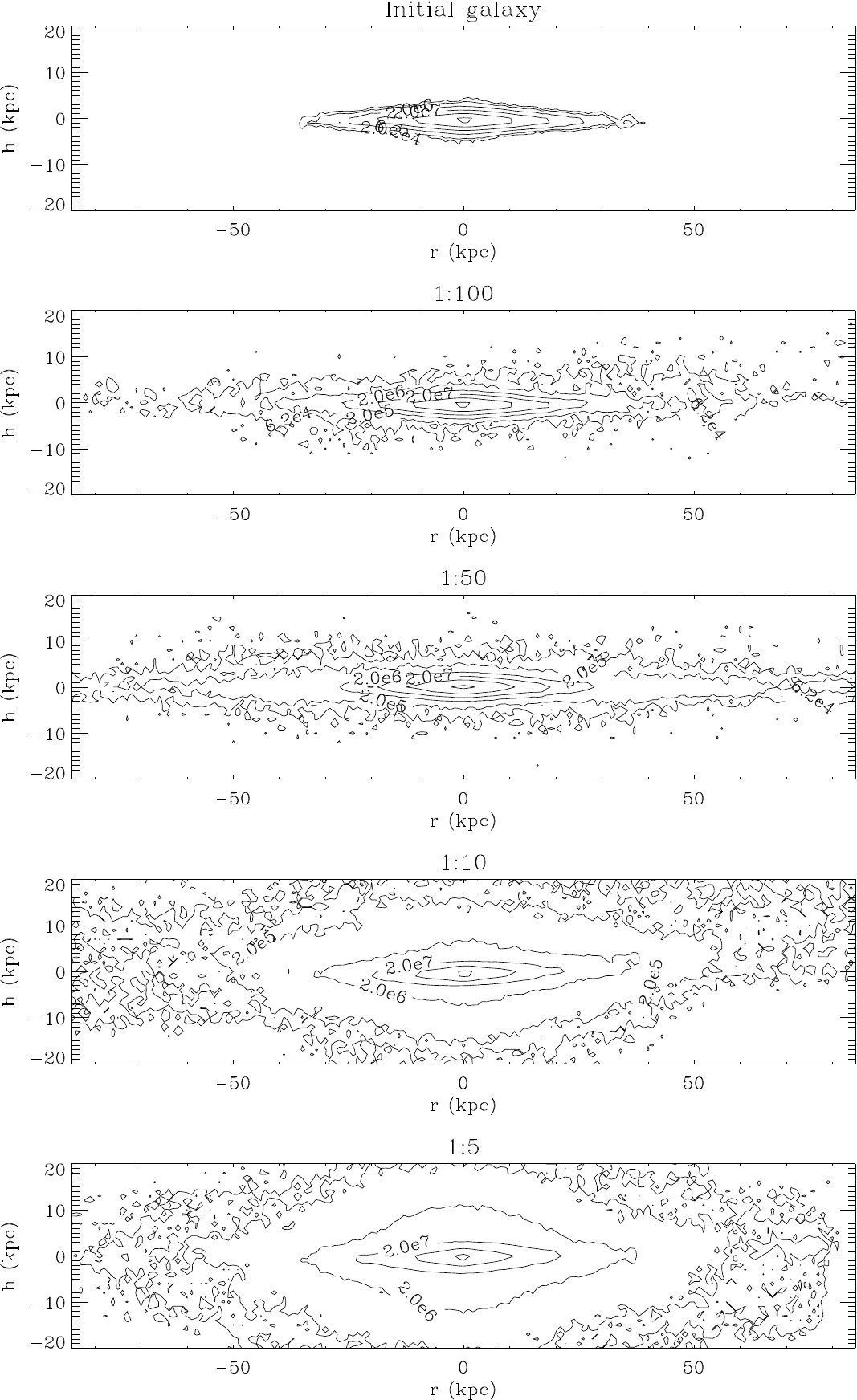}}
\caption{The resulting edge-on density distribution of
the disc-component of the remnant after a relaxation time of
${\sim}\,3.5~\mathrm{Gyr}$ for the Family~2 mergers
with $\Omega=0\dg$ and $\alpha=0\dg$.
For comparison, the uppermost panel shows the undisturbed disc evolved
for the same time as the merger simulation.
The other panels show, from top to bottom, the results for the
\mbox{$\mu=1:100$, $1:50$, $1:10$, and $1:5$} merger mass ratio
simulations.
The contour lines are in units of $\Msun/\mathrm{kpc}^2$.}
\label{fig:discgrowth}
\end{figure}

As shown in Section~\ref{sec:Results}, the host disc galaxy loses its
initial form and becomes more and more isotropic for minor mergers
in general and for larger satellite masses in particular.
However, we find that the appearance of shell or stream structures
does not depend, in general, on the merger mass ratio but is closely
correlated with the impact angle $\alpha$:
for small $\alpha$ values up to $\alpha\sim10\dg$, all mergers
produce shells but no streams, while for larger values of $\alpha$
only streams appear, and no shells at all.
Figure~\ref{fig:shell} illustrates this behavior:
the left panels show the face-on (upper panel) and edge-on (lower
panel) projections of the mass distribution of a $\mu=1:100$ satellite
galaxy merger with $\alpha=10\dg$, $\Omega=0\dg$ and $\omega=30\dg$,
while the right panels show the same for the identical merger scenario
but $\alpha=40\dg$.
As can be seen very clearly, the merger with the small impact angle
produces only shell structures, while the merger with the large impact
angle causes streams but no shells.
This is in good agreement with the results from \citet{Amorisco2015}
who also reported that shells are made by mergers on radial orbits,
while streams appear for more circular orbits.
This behavior can be seen in more detail in Figs.~\ref{fig:fam1top}
and~\ref{fig:fam1side} in Appendix~\ref{app:b}, where the same diagrams
as in Fig.~\ref{fig:shell} are shown for the full sample of simulations
from Family~1.

We also find that there is no effect of the orbital angle $\Omega$
on the appearance of shells or streams, although all structures are
slightly less pronounced for the face-on ($\Omega=90\dg$) collisions
(see Figs.~\ref{fig:fam2top} and~\ref{fig:fam2side}).
To summarize, we find clear evidence that galaxy shells only occur
for central (radial) hits, and are more prominent for orbits along
the major axis of the host and less pronounced for face-on mergers.
By visual inspection of the formation of streams and 
shells in the simulations, we find that the lifetime of streams appears 
to be longer than that of shells. The latter seem to be more rapidly 
smoothed out into a continuous distribution, as it can be seen in 
Fig.~\ref{fig:shell}.

Under certain conditions it is possible that the mergers mainly
increase the host disc size without puffing up the disc.
Figure~\ref{fig:discgrowth} shows the edge-on view of the host disc
for the Family~2 mergers with $\Omega=0\dg$ and $\alpha=0\dg$, with
$\mu=1:5$, $1:10$, $1:50$, and $1:100$ from bottom to top.
The uppermost panel shows the undisturbed disc galaxy evolved for
$t=3.5~\mathrm{Gyr}$.
As can be clearly seen, the disc of the galaxy grows up to two times
its original size by a $\mu=1:100$ merger, and even more for the
$\mu=1:50$ merger.
Meanwhile, the disc-height remains constant with a scale-height of
$h\approx 1.5~\mathrm{kpc}$ for both mini mergers.
With increasing mass of the merging satellite, the size of the disc
does not increase any further, but instead the disc height starts to
grow significantly while the disc structure is destroyed.
In the lower panels of Fig.~\ref{fig:discgrowth} this effect can be
seen very prominently.

Besides the mass ratio, the orbital parameters are also very important
in forming such large discs.
An increase in disc size can only be seen for $\Omega < 30\dg$.
If the inclination of the merger rises to $\Omega \geq 30\dg$, the
density distribution becomes much more spherical.

The observation of such a giant, low-surface-brightness stellar disc
around NGC~2841 has just recently been reported by \citet{Zhang2018}.
The observed disc has a radial extent of ${\sim}\,70~\mathrm{kpc}$ 
with a mass surface density at the edge of ${\sim}0.1~\Msun/pc^2$ 
which is similar to the outermost contour lines in Fig.~\ref{fig:discgrowth}. 
Nevertheless, as we are assuming a constant mass-to-light ratio, a 
proper comparison would require mimicking the observations (e.g. 
computing the corresponding fluxes in the corresponding band and 
applying surface brightness cuts). This suggests mini 
mergers as a possible explanation for the origin
of this extended disc, in particular also because NGC~2841 at the
same time does not show signs of a spherical stellar halo.
It is quite possible that more than one mini merger would be
needed to build up the extended disc to the observed surface
brightness; however, as there are now several observational reports
of satellite galaxies being aligned in planes around their host
galaxies \citep[e.g.,][]{Tully2015,Mueller2016,Fritz2018}, multiple
mini mergers building up such a disc would indeed appear possible,
although we cannot yet comment on the probability of this happening.

\section{Summary and conclusion}
\label{sec:Discussion}

Mergers play an essential role in a galaxy's evolution, shaping both
its morphology and driving its mass growth.
Numerical simulations of mergers have become a key tool in
understanding the diversity of observed galactic structures and
their environments.
Cosmological simulations, having now reached resolution levels making
it possible to resolve individual galaxies even in huge simulation
volumes with box lengths of a gigaparsec or more, have proved important
in explaining the statistical properties of galaxy populations.
For understanding the details of merger processes, however,
individually set up high-resolution merger simulations with carefully
controlled configurations are still indispensable.

We performed 36 simulations of isolated mergers between two disc
galaxies with mass ratios varying between $\mu= 1:5$ and $1:100$
with high particle resolution.
With these simulations, we provide a systematic study of the influence
of different orbit configurations on the mass distribution of the
final merger remnant and specifically the mass distribution of the
disrupted satellite galaxy within the final merger remnant.

We find that mergers with mass ratios from $\mu= 1:10$ to $1:5$,
so-called minor mergers, already have a sufficient impact to
significantly alter the morphological appearance of the host galaxy,
as measured by the S\'ersic-index.
Mergers with smaller satellites (mini mergers), however, barely change
the S\'ersic-index of the host, leading primarily to an increase of
the disc size and a slight puffing-up of the disc, depending on the
infall direction of the satellite relative to the host disc.

Perhaps the most important result is that especially the mini mergers
deposit their mass at much larger radii than expected from assuming
that the breakup of the satellite results purely from the tidal
forces of the host galaxy -- the medians of the radial post-merger
distribution of the former satellite stars in the simulations are
larger by up to a factor of~2 in radius than predicted by this model.
However, if we do not assume that the satellite core is free-falling
with the stripped particles (because it will be subject to additional
forces such as dynamical friction that do not affect the stripped
particles) and instead consider only the actual binding forces,
then the predicted medians agree much better with those from the
simulations.
This indicates that a description based solely on tidal forces does
not capture all essential aspects of actual mergers.

While minor mergers always deposit part of their mass to the center of
the host independent of the orbital parameters, mini mergers cannot
reach the center if they approach the host on circular orbits close
to the disc plane of the host.
In fact, mini mergers can contribute to building up a stellar halo
around the host without significantly disturbing the center, depositing
most of their mass in the outskirts.
If the satellite approaches in the disc plane, this halo can even
be disc-like in shape, thus offering a possible explanation for the
observed existence of such extended halo-less discs \citep{Zhang2018}.

Interestingly, the impact parameter angle $\alpha$ and the orbital
plane angle $\Omega$ have opposing effects on the radial mass
distribution of the satellite particles within the remnant, while
an inversion of the internal rotation of the merger only has a
minor effect.
Increasing $\alpha$ leads to a decrease of the satellite contribution
to the center of the merger remnant, while increasing $\Omega$ causes
more satellite particles to be deposited at the center.
Furthermore, $\alpha$ plays a crucial role for the emergence of shell
and stream features:
streams appear for $\alpha\ge30\dg$ (pericentric distance
$\ge15~\mathrm{kpc}$), and values of $\alpha\le10\dg$ (pericentric
distance $\le1.8~\mathrm{kpc}$) lead to distinct shell structures,
independent of the choice of $\Omega$.
Intermediate values of $\alpha$ lead to structures showing features
of both streams and shells.
To conclude, streams are a strong indication of nearly circular infall
of a satellite (with a large angular momentum), whereas the appearance
of shells clearly points to (nearly) radial satellite infall.

\section*{Acknowledgements}
We would like to thank the reviewer for his/her insightful comments and constructive remarks.
The Simulations were run at the Leibniz Computing Center, Germany,
as part of the project pr86re.
UPS and BPM acknowledge an Emmy Noether grant funded by the Deutsche
Forschungsgemeinschaft (DFG, German Research Foundation) with the
project number MO 2979/1-1.


\balance
\bibliographystyle{aa}
\bibliography{sources}

\onecolumn
\appendix

\section{Shells and Streams for all Mergers}
\label{app:b}

We show the stellar particle distributions of the satellite galaxy at
a time $t\sim3.5~\mathrm{Gyr}$ after the merger event (i.e., 4.2~Gyr
after the beginning of the simulation, with the merger event taking
place approximately 0.7~Gyr after the simulation begins, this value
varying slightly with the impact angle $\alpha$).
All distributions are shown in both edge-on ($X$--$Z$) and face-on
($X$--$Y$) views, referring to the orientation of the host galaxy
(which, however, is not shown in these plots).
In all figures, colors show the stars from the different components
of the initial satellite galaxies:
black---disc stars; red---bulge stars; yellow---gas and newly formed
stars, as in Fig~\ref{fig:shell}.
\vspace{2\baselineskip}

\begin{figure}
\here
\parbox[t]{240pt}{
\centerline{\includegraphics[scale=0.54]{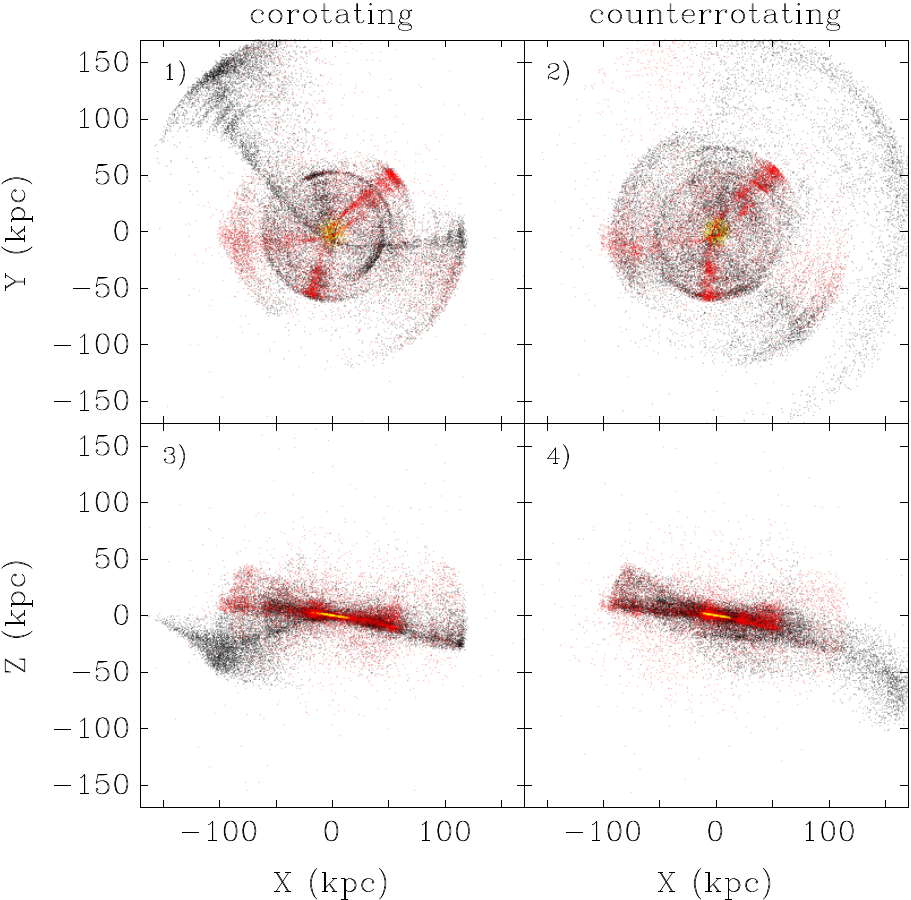}}
\caption{Face-on (top panels) and edge-on (bottom panels) views of
the distribution of the satellite particles for the special case
simulations with $\alpha=10\dg$ and $\Omega=10\dg$, where the discs
are corotating (left panels) and counterrotating (right panels).}
\label{fig:hellspecial2}
}\hfill
\parbox[t]{240pt}{
\centerline{\includegraphics[scale=0.54]{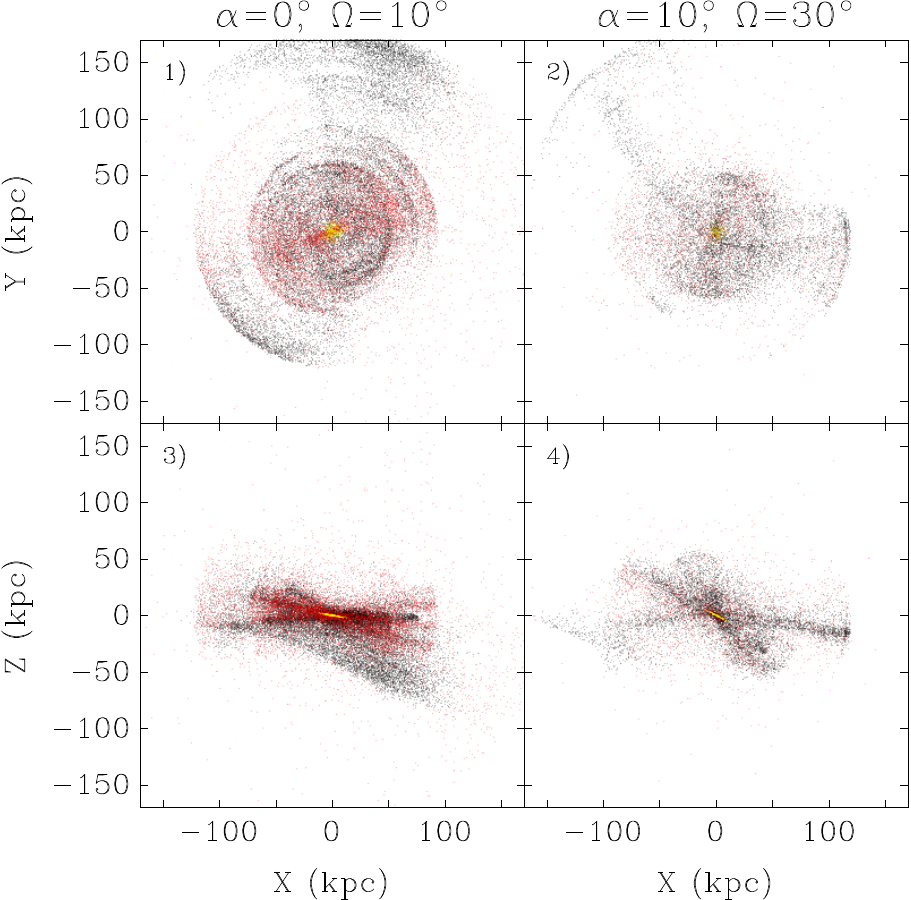}}
\caption{Face-on (top panels) and edge-on (bottom panels) views of
the distribution of the satellite particles for the special case
simulations with $\alpha=0\dg$, $\Omega=10\dg$ (left panels), and
$\alpha=10\dg$, $\Omega=30\dg$ (right panels).}
\label{fig:hellspecial}
}
\end{figure}

\begin{figure*}
\centerline{\includegraphics[scale=0.54]{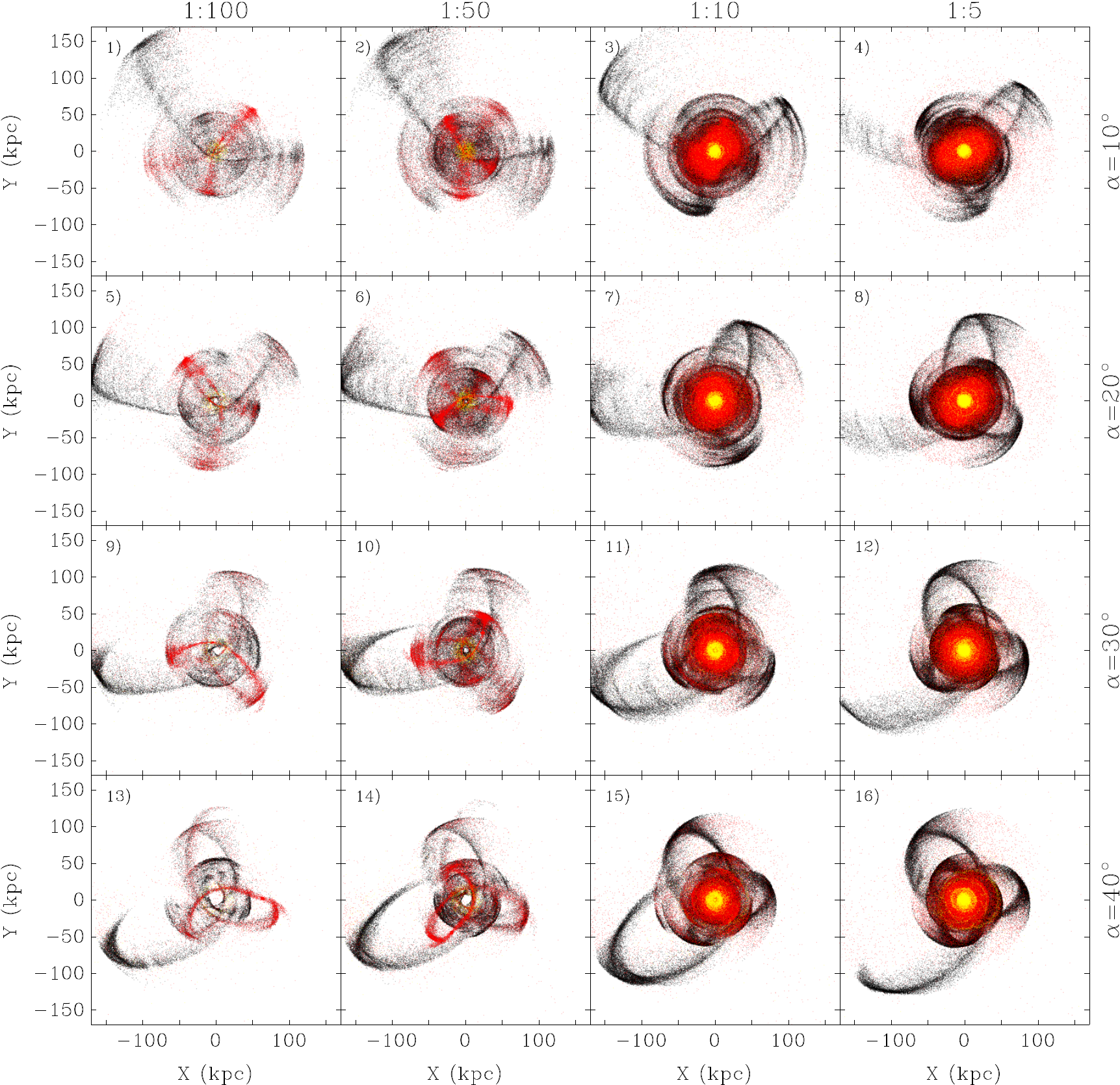}}
\caption{Face-on view of all satellite particles for all Family~1
mergers.}
\label{fig:fam1top}
\end{figure*}

\begin{figure*}
\centerline{\includegraphics[scale=0.54]{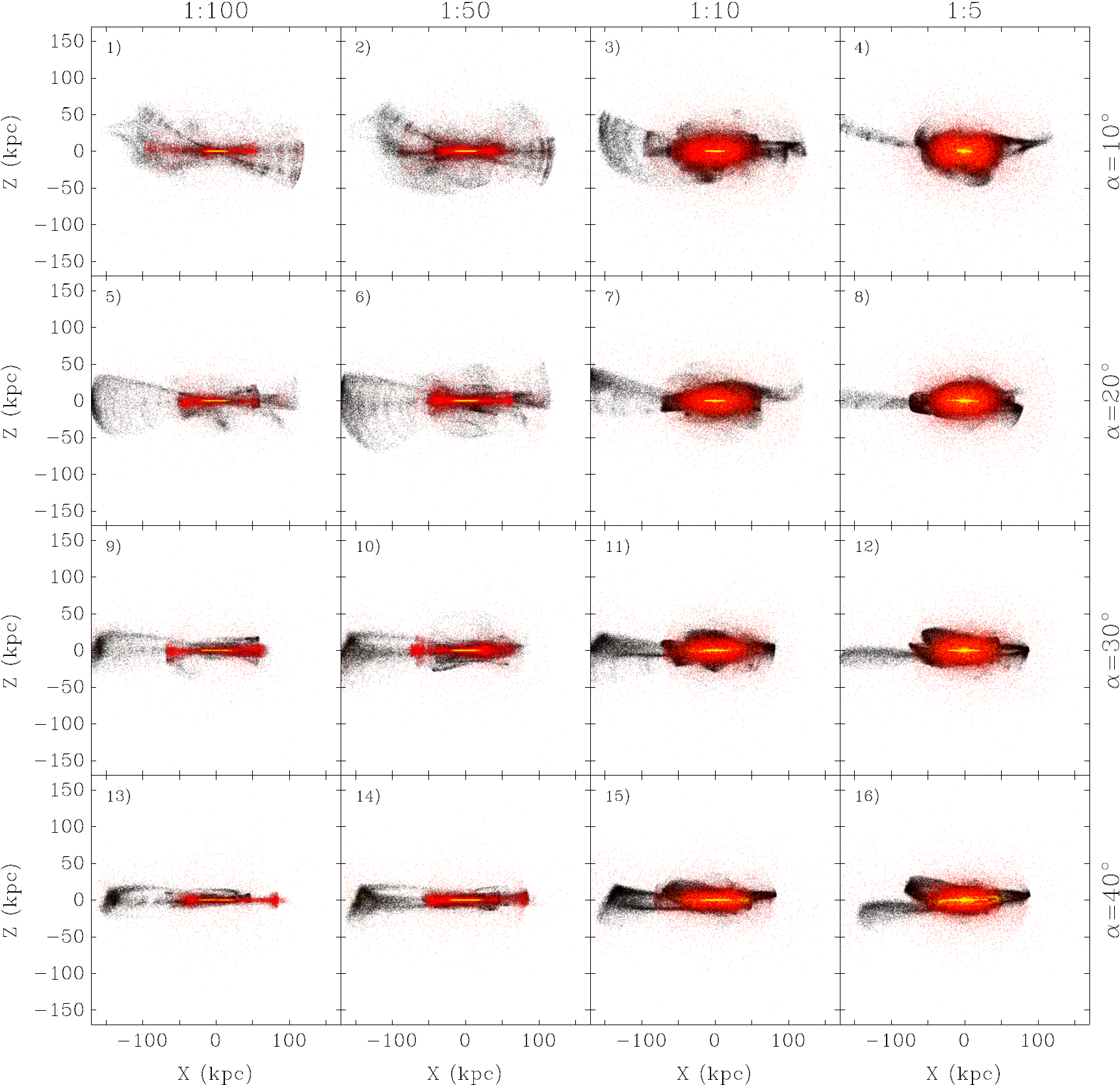}}
\caption{Edge-on view of all satellite particles for all Family~1
mergers.}
\label{fig:fam1side}
\end{figure*}

\begin{figure*}
\centerline{\includegraphics[scale=0.54]{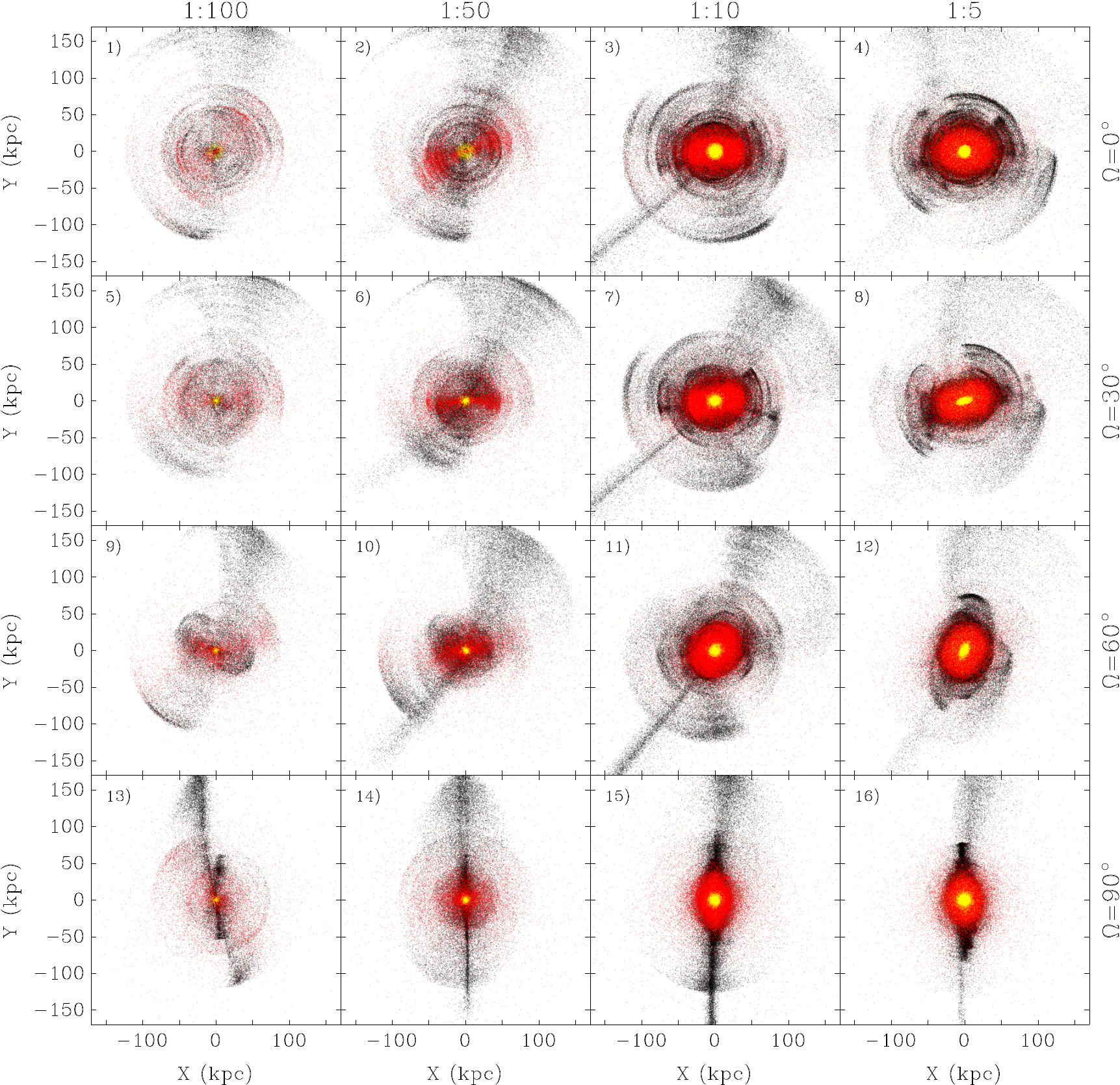}}
\caption{Face-on view of all satellite particles for all Family~2
mergers.}
\label{fig:fam2top}
\end{figure*}

\begin{figure*}
\centerline{\includegraphics[scale=0.54]{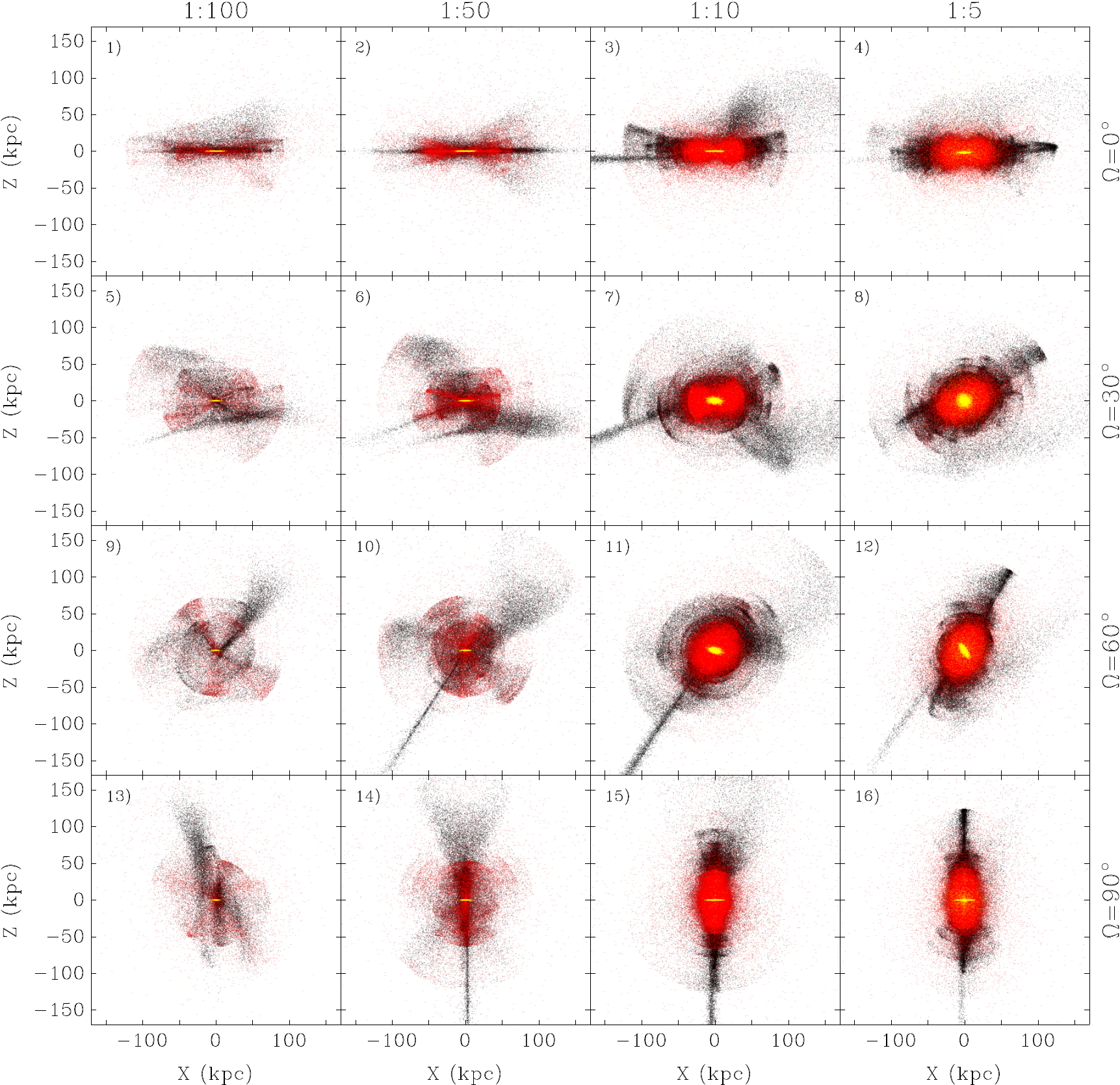}}
\caption{Edge-on view of all satellite particles for all Family~2
mergers.}
\label{fig:fam2side}
\end{figure*}


\label{lastpage}
\end{document}